# Cyber Attack Mitigation Framework for Denial of Service (DoS) Attacks in Fog Computing

**Fizza Khurshid[1], Umara Noor[2], Zahid Rashid[3]**

[1]Department of Computer Science, Faculty of Computing and Information Technology, International Islamic University, Islamabad, Pakistan
[2]Department of Software Engineering, Faculty of Computing and Information Technology, International Islamic University, Islamabad, Pakistan
[3]Technology Management Economics and Policy Program, College of Engineering, Seoul National University, 1 Gwanak-Ro, Gwanak-Gu, 08826 Seoul, South Korea
fizza.mscs1105@iiu.edu.pk, umara.zahid@iiu.edu.pk, rashidzahid@snu.ac.kr

**Abstract**

Innovative solutions to cyber security issues are shaped by the ever-changing landscape of cyber threats. Automating the mitigation of these threats can be achieved through a new methodology that addresses the domain of mitigation automation, which is often overlooked. This literature overview emphasizes the lack of scholarly work focusing specifically on automated cyber threat mitigation, particularly in addressing challenges beyond detection. The proposed methodology comprise of the development of an automatic cyber threat mitigation framework tailored for Distributed Denial-of-Service (DDoS) attacks. This framework adopts a multi-layer security approach, utilizing smart devices at the device layer, and leveraging fog network and cloud computing layers for deeper understanding and technological adaptability. Initially, firewall rule-based packet inspection is conducted on simulated attack traffic to filter out DoS packets, forwarding legitimate packets to the fog. The methodology emphasizes the integration of fog detection through statistical and behavioral analysis, specification-based detection, and deep packet inspection, resulting in a comprehensive cyber protection system. Furthermore, cloud-level inspection is performed to confirm and mitigate attacks using firewalls, enhancing strategic defense and increasing robustness against cyber threats. These enhancements enhance understanding of the research framework's practical implementation and assessment strategies, substantiating its importance in addressing current cyber security challenges and shaping future automation mitigation approaches.

## 1. Introduction

The internet has provided humanity with several benefits, including access to a vast amount of knowledge and the ability to connect with businesses, friends, and families around the globe. The digital revolution has transformed our way of life and has become a vital component of our daily routine [1]. However, with growing technologies, there is no safeguard of our information and cyber-crimes are becoming common. Cyber threats are increasing rapidly, 94% of malware is delivered via email. There are over about 2,244 attacks that happen daily on the internet which break down to nearly 1 cyber-attack every 39 seconds. The situation is getting worse as in 2021, 82.6% of organizations located world-wide experienced cyber-attacks. It is predicted that in 2025 the cost of cyber-attack damage will be 10.5 trillion USD [2]. The average annual loss faced by organizations is estimated as 79,841 USD [3]. The inability to mitigate these cyber threats can result in loss of business productivity, reputation, and finances [3]. This financial toll extends to indirect costs associated with recovery and system restoration in addition to direct losses. These cyber-threats cannot be ignored since consequences extend beyond the financial realm. Such an attack targets companies with varied effects from reputation loss, falling revenues and decrease corrective measurements taken. But while cyberspace becomes even more complex, the need for devices that fight cyber threats grows. Given that the business environment is becoming increasingly interdependent and vulnerable to internet platforms; a firm may face extinction unless it installs strong cyber safety infrastructure systems that infringe on data security standards also, and as is it shown in the figure 1 that DDoS attack have constantly increasing per year, it reaches 15.4 million in the year 2023 [4]. To overcome the effect of these cyber threats, cyber threat mitigation strategies need to be implemented in a timely fashion.

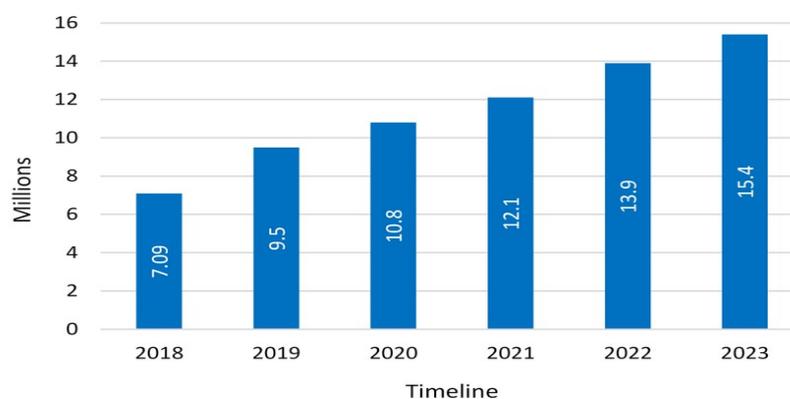

*Figure 1: Global DDoS statistics*

## 1.1 Understanding Cyber Threats

In today's interconnected digital landscape, cyber threats pose a significant risk to the security and stability of networked systems. As organizations increasingly rely on digital technologies for their operations, they become more vulnerable to various forms of cyber threats. A cyber threat refers to any malicious act or activity that seeks to compromise the confidentiality, integrity, or availability of digital information or systems. These threats can manifest in numerous ways, from common malware infections and phishing attacks to sophisticated cyber espionage campaigns and ransomware outbreaks [5].

The impact of cyber threats extends beyond mere inconvenience; it can be devastating, resulting in substantial financial losses, reputational damage, and disruption of critical services. For businesses, the ramifications of a successful cyber-attack can be catastrophic, leading to loss of revenue, legal liabilities, and erosion of customer trust. Similarly, for individuals, cyber threats can jeopardize personal privacy, financial security, and even physical safety in certain cases. Therefore, understanding the scope and impact of cyber threats is essential for implementing effective cyber security measures and safeguarding against potential risks [6].

## 1.2 Navigating Distributed Denial of Service (DDoS) Attacks

One prominent manifestation of cyber threats is Distributed Denial of Service (DDoS) attacks. These attacks leverage the distributed power of numerous compromised devices to orchestrate a coordinated assault on a target's network infrastructure. Malicious actors exploit vulnerabilities in these devices, often referred to as "bots" or "zombies," to hijack them and turn them into unwitting participants in the attack. Once under the attacker's control, these compromised devices are directed to generate and send a massive volume of traffic to the target, overwhelming its network resources and rendering it inaccessible to legitimate users.

DDoS attacks can take various forms, including volumetric attacks, protocol attacks, and application layer attacks. Volumetric attacks aim to flood the target with a deluge of traffic, consuming its bandwidth and network resources. Protocol attacks exploit weaknesses in network protocols, such as TCP, UDP, and ICMP, to disrupt communication between networked devices. Application layer attacks target specific applications or services, exploiting vulnerabilities in their software or infrastructure to disrupt their functionality.

The impact of DDoS attacks can be severe, leading to downtime, financial losses, and reputational damage for the target organization. In some cases, DDoS attacks may also serve as a smokescreen for more nefarious activities, such as data theft or network infiltration. Therefore, understanding the nature and tactics of DDoS attacks is critical for organizations to develop robust defense mechanisms and mitigate their impact effectively [7].

A TCP SYN flood is one of Distributed Denial of Service (DDoS) attacks that expose the flaw mechanism in Transmission Control Protocol's three way handshake process. In this attack, rogue agents swamp a targeted node with SYN requests in an effort to exhaust the resources of that system as shown below in figure 2 [8]. Unlike a normal three-way handshake that tells the connection establishment is over through an ACK packet, this attacker does not end with sending back such an ACK. As a result, the target system expends resources for every incoming SYN request and wait in vain by hours on end to receive confirmation that never arrives. This results into the depletion of resources such as memory and processing capacity hence making the system incapable of handling legitimate connection requests. A TCP SYN flood attack has a great effect, usually leading to the loss of service or even system failure altogether. Mitigation measures include the rate limiting aimed at controlling incoming connection rates, SYN cookies for validation and installing intrusion prevention systems as well as firewalls which filter malicious traffic in order to prevent DDoS attacks.

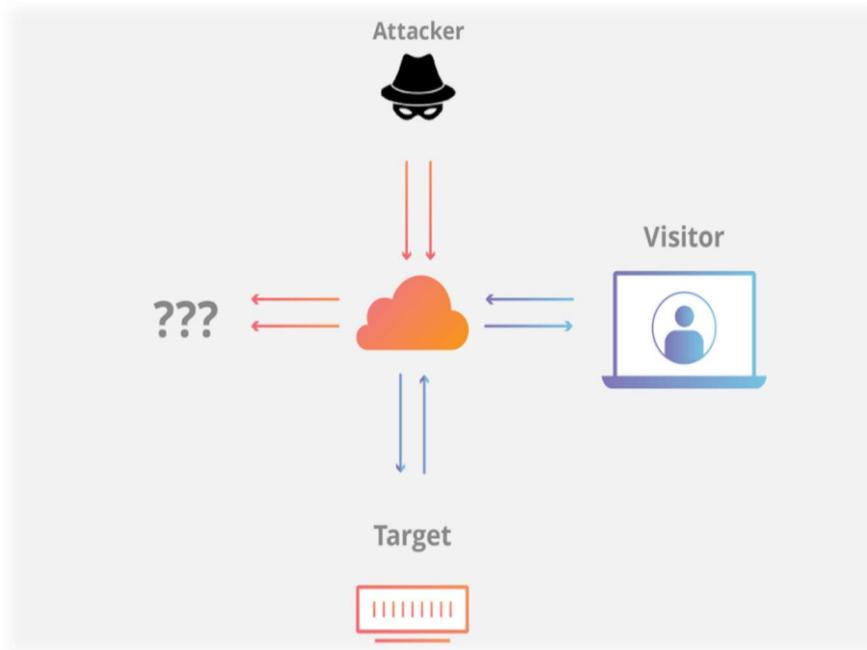

*Figure 2: TCP SYN Flood Attack*

## 1.3 Unveiling the Potential of Fog Computing in Cyber Security

In response to the escalating threat landscape, Fog computing has emerged as a promising paradigm for addressing cybersecurity challenges, including DDoS attacks. Fog computing extends the capabilities of traditional cloud computing by decentralizing data processing and storage closer to the edge of the network, thereby reducing latency and enabling real-time analysis of network traffic [9]. Unlike cloud computing, which centralizes data processing in remote data centers, fog computing distributes computational resources across a distributed network of edge devices, such as routers, switches, and IoT devices [9].

Fog nodes, also known as fog devices or fog servers, play a pivotal role in the fog computing architecture. These nodes are strategically deployed at the network's edge, where they serve as points of data processing, storage, and analysis. By distributing computational resources across fog nodes and cloud servers, fog computing enhances the responsiveness and scalability of cyber security defenses, making it particularly effective in mitigating DDoS attacks and other cyber threats. Figure 3 illustrates the fog computing architecture.

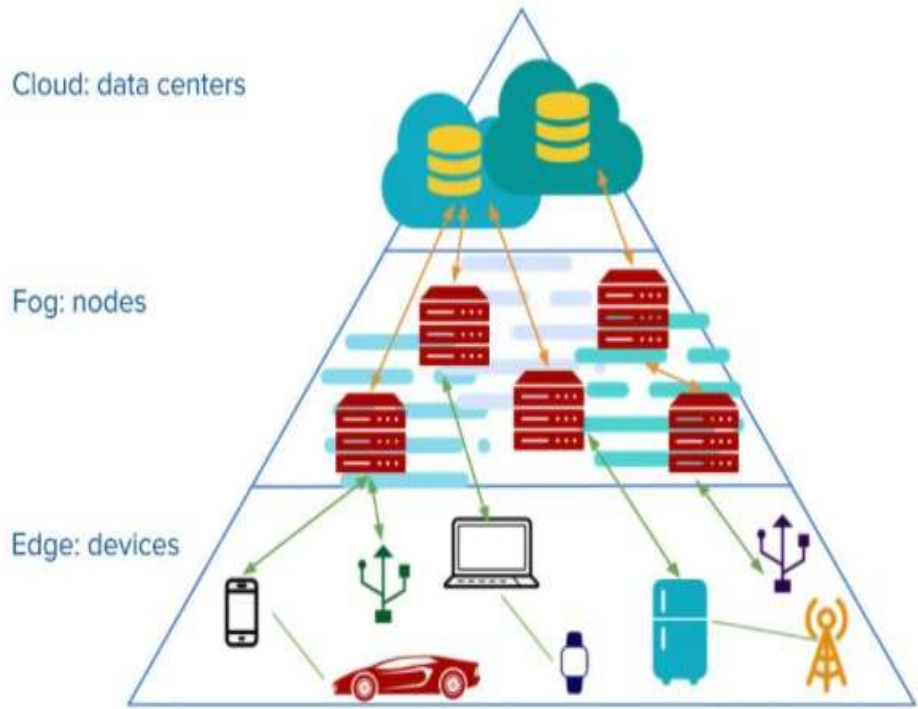

*Figure 3: Fog computing architecture*

Fog computing offers several advantages over traditional cloud computing models, including reduced latency, improved reliability, and enhanced privacy and security. By processing data closer to the source, fog computing minimizes the time it takes for data to travel between devices and data centers, enabling faster response times and real-time analysis of network traffic. Additionally, fog computing allows organizations to maintain greater control over their data and infrastructure, reducing the risk of data breaches and unauthorized access [10].

### 1.4 The Role of Internet of Things (IoT) in Fog Computing

The Internet of Things (IoT) plays a crucial role in the Fog computing ecosystem, as it comprises a vast network of interconnected devices that generate and exchange data. These devices, ranging from sensors and actuators to smart appliances and industrial machinery, contribute to the continuous stream of data that fog computing systems analyze for potential security threats. By harnessing the power of IoT devices, Fog computing enables organizations to collect and analyze data in real-time, detect anomalies, and respond promptly to security incidents [5].

IoT devices generate a wealth of data that can be leveraged to enhance cyber security defenses and mitigate the risk of DDoS attacks and other cyber threats [6]. For example, sensors deployed in critical infrastructure, such as power plants or transportation systems, can detect anomalies or unusual patterns of activity that may indicate a cyber-attack in progress. By analyzing this data in real-time, organizations can identify and respond to security incidents more effectively, reducing the impact of cyber threats on their operations.

Fog computing extends the capabilities of traditional cloud computing by decentralizing data processing and storage closer to the edge of the network, thereby reducing latency and enabling real-time analysis of network traffic. This distributed computing paradigm plays a pivotal role in mitigating DDoS attacks and other cyber threats, as it provides a more secure and responsive environment for processing data generated by IoT devices.

## 1.5 Empowering Fog Computing with Network Function Virtualization (NFV)

Network Function Virtualization (NFV) is a technology framework and architectural approach that aims to virtualize and consolidate networking functions traditionally performed by dedicated hardware appliances. In NFV, these network functions, such as firewalls, routers, load balancers, and intrusion detection systems, are decoupled from proprietary hardware and implemented as software-based services that can run on standard commercial off-the-shelf (COTS) hardware. This allows network operators and service providers to deploy and manage network services more flexibly, efficiently, and cost-effectively by leveraging virtualization and cloud computing technologies [11].

NFV enables dynamic provisioning, scaling, and chaining of network functions to meet changing demand and optimize resource utilization, leading to greater agility, scalability, and cost savings in network infrastructure deployment and operation. By abstracting network functions from underlying hardware, NFV simplifies network management and reduces operational complexity, enabling organizations to rapidly deploy new services, automate network operations, and adapt to evolving business and technology requirements [12]. Additionally, NFV promotes interoperability and vendor-agnosticism by standardizing interfaces and protocols, fostering innovation and competition in the network services ecosystem. Overall, NFV plays a crucial role in modernizing and transforming network infrastructure to meet the demands of increasingly dynamic and data-intensive digital environments.

## 1.6 Fortifying Defenses with Firewalls

Firewalls serve as crucial components of cyber security defenses, acting as the first line of defense against unauthorized access and malicious traffic. In the context of DDoS mitigation, firewalls play a critical role in filtering incoming traffic based on predefined rules and policies. Rule-based packet filtering, a common firewall technique, involves inspecting each packet against a set of rules and allowing or blocking traffic based on predefined criteria. Deployed at network boundaries and internal segments, firewalls enforce security policies and protect against DDoS attacks and other cyber threats [13].

Firewalls come in various forms, including traditional perimeter firewalls, internal firewalls, and next-generation firewalls, each offering different levels of protection and functionality. Perimeter firewalls are deployed at the network perimeter, where they filter traffic entering and exiting the network, while internal firewalls segment internal network traffic to prevent lateral movement by attackers. Next-generation firewalls incorporate advanced features, such as intrusion detection and prevention, application control, and SSL inspection, to provide more comprehensive protection against modern cyber threats [14].

## 1.7 Understanding Rule-Based Packet Filtering for Cyber Defense

Packet filtering is a fundamental technique used in network security to inspect and selectively forward or discard packets based on specified criteria [15]. Employed in firewalls, routers, intrusion detection systems, and other network security devices, packet filtering techniques enforce security policies and protect against cyber threats, including DDoS attacks. Techniques include rule-based packet filtering, stateful packet inspection, deep packet inspection, and application-layer filtering, among others. Each technique has its strengths and limitations, and organizations may deploy multiple techniques in combination to enhance their cybersecurity defenses and mitigate the risk of DDoS attacks and other cyber threats [16].

Packet filtering techniques enable organizations to analyze network traffic and selectively allow or block packets based on predefined criteria. Rule-based packet filtering examines each packet against a set of rules and policies, allowing organizations to control traffic flow based on its characteristics and origin [17]. Stateful packet inspection maintains a stateful connection table to track the state of network connections, enabling more sophisticated filtering based on connection state and context [18]. Deep packet inspection analyzes packet contents at the application layer, allowing organizations to inspect and filter traffic based on application-specific protocols and payloads. Application-layer filtering focuses on filtering traffic based on application-layer protocols and attributes, providing more granular control over network traffic [15].

Rule-based packet filtering is a fundamental technique used in firewalls and other network security devices to control traffic flow based on predefined rules and policies. Each packet traversing the network is inspected against a set of rules specifying criteria such as source and destination IP addresses, ports, protocols, and packet attributes. Based on the match between the packet and the rules, the firewall either allows the packet to pass through, drops it, or forwards it to a different destination. Rule-based packet filtering enables organizations to enforce security policies, block malicious traffic, and protect against DDoS attacks and other cyber threats effectively [17].

Rule-based packet filtering provides organizations with granular control over network traffic, allowing them to define specific rules and policies tailored to their security requirements. These rules can be based on various criteria, including IP addresses, port numbers, protocols, and packet attributes, enabling organizations to filter traffic based on its characteristics and origin. Additionally, rule-based packet filtering can be combined with other security measures, such as intrusion detection and prevention systems, to provide layered protection against cyber threats.

## 1.8 Proactive Measures and Mitigation Strategies

Mitigation in cyber security refers to the comprehensive set of proactive measures and strategies designed to prevent, detect, and mitigate the impact of cyber threats on an organization's network infrastructure and digital assets [19], [20]. These measures are essential for safeguarding against various types of attacks, including Distributed Denial of Service (DDoS) attacks and Denial of Service (DoS) attacks. Mitigation strategies encompass a diverse range of techniques, beginning with the continuous monitoring and analysis of network traffic patterns to identify anomalies or suspicious activities that may indicate an ongoing attack. Anomaly detection systems, often powered by machine learning algorithms and statistical analysis, play a crucial role in this process by flagging deviations from normal behavior and triggering alerts for further investigation. Additionally, organizations deploy rate-limiting mechanisms to control the flow of incoming and outgoing traffic, thereby preventing network congestion and mitigating the impact of volumetric attacks, such as DDoS and DoS attacks. Furthermore, traffic diversion techniques are employed to redirect malicious traffic away from targeted systems or services to alternative resources or network segments, minimizing the disruption caused by an attack. Access control lists (ACLs) and Intrusion Prevention Systems (IPS) are also utilized to filter and block traffic from known malicious sources, while cloud-based DDoS protection services offer scalable and resilient defense mechanisms to filter and scrub malicious traffic before it reaches the organization's network infrastructure. By implementing a combination of these mitigation strategies, organizations can enhance their cyber security posture, mitigate the risk of successful cyber-attacks, and maintain the resilience and reliability of their network infrastructure in the face of evolving threats [21].

As the emerging trends in cyber security literature indicate, automation of threat detection has received a great deal of attention recently and is supported by numerous researches [22], [23], [24]. But there is a significant research gap in the area of automated cyber threat mitigation. This anomaly is due to the lack of a comprehensive taxonomy that enumerates mitigations for each stage of the cyber kill chain model. The cyber kill chain is a cyber-security model developed by Lockheed Martin that tracks the stages of a cyber-attack, detects vulnerabilities, and assists security teams in preventing attacks at each point of the chain [25]. The term was originally used by the military to describe the organization of an attack. It involves finding a target, dispatching the target, making a choice, issuing an order, and eventually destroying the target [25]. Such taxonomy would be necessary in allowing security professionals to automate the mitigating process effectively. While there are various taxonomies around the literature on cyber threat detection, a critical need remains in this area of studies for an extensive and sufficiently inclusive that covers all relevant aspects related to identification stages within such processes as cyber kill chain development/plotting etc. However, this gap is a serious challenge towards developing strategic and effective solutions to counter various types of cyber-attacks. However, the absence of investigations into automated cyber threat mitigation methods complicates matters even more. While there are many automated ways to detect cyber threats, researchers have almost entirely neglected the presentation of automated methods for mitigating those same threats. This suggests the importance of categorizing cyber threat mitigation approaches as automatic or manual methods based on what corrective actions should be taken.

The current security control procedures are an integral part of automated cyber threat mitigation. Data confidentiality, reliability and integrity protection mechanisms employed by organizations are security measures such as firewalls, robust authentication procedures setting up timely deployment of security patches in their network. As cyber security evolves, efforts at research and development should aim to narrow the integration gap between automated detection software solutions and mitigation systems. For a robust cyber security framework, both factors need to be well understood and implemented to eliminate the persistent dynamic nature, which is referred to as the cyberspace threat environment.

This study can be regarded as a revolutionary step in this sphere due to the holistic taxonomy for threat mitigation cyber security that has profound impacts on the global level. The main purpose of such taxonomy is to specify which mitigation actions can be automated from the point of view of the security control mechanisms that are constantly in use. The research may be a source for corporations that aim to enhance their cyber security posture by automating threat response processes in an efficient way through such classification of tactics.

Apart from the taxonomy, this research proposes a novel contribution: a system of automatization to prevent Distributed Denial of Service (DDoS) attacks based on TCP-SYN flooding, which has been introduced by the TCP protocol for fog computing. This framework is a new solution to the problem of dynamic obstacles generated by DDoS attacks with TCP-SYN flooding in the fog environment. This automated mitigation method is important because it is active in detecting and stopping DDoS attacks that use TCP-SYN flooding; one of the most common cyber threats. However, its implementation within the fog computing paradigm provides flexibility in response to dispersed computation attributes, thus ensuring a more sophisticated and powerful resistance against such attacks. The complete dedication of the research to offering practical solutions in addition to conceptual innovation is further emphasized by combining automated TCP-SYN flooding DDoS attack mitigation framework and cyber threat mitigating taxonomy. This two-pronged approach advances the cyber security through provision of information and resources to help organizations navigate adequately, effectively & resiliently around such complexities in today's threat landscape.

### 1.9 Research Gaps/ Limitations

In the existing literature, the limitations are:

Currently, a complete taxonomy related to mitigations of all kinds of cyber threats is not available. [26]

Previous research has not undertaken the classification of cyber threat mitigation strategies into automated and non-automated categories within the taxonomy. [26]

Local firewall activation by updating firewall rules (TCP-SYN DdoS attack) [27]

SDN flow rules-based mitigation. [28]

Reconfiguration [29]

Application-aware firewall [30]

Hardware upgradation. [31]

### 1.10   Problem Statement

A comprehensive cyber threat mitigation taxonomy is not available that can be helpful in effectively mitigating the data breaches automatically.

There is need to automate the local firewall activation mitigation for TCP-SYN flooding DDoS attack that is performed manually in the existing literature.

### 1.11   Research Questions

1. What are the state-of-the-art of cyber threat mitigation strategies?
2. How the cyber threat mitigation taxonomy will be created?
    a. What sources will be considered for collecting the mitigations data reported so far?
3. Which of the cyber threat mitigations can be automated using existing security controls?
4. How the proposed automated threat mitigation framework will be implemented?

### 1.12   Aim and Objectives

**Aim**

- ➢ To create a comprehensive threat mitigation taxonomy.
- ➢ To develop an automated cyber threat mitigation framework for DoS attack.

**Objectives**

- ➢ To create a comprehensive cyber threat mitigation taxonomy using the state-of-the-art mitigation strategies.
- ➢ Classify which mitigation techniques can be automated and non-automated.
- ➢ Develop an automated threat mitigation framework for Denial-of-Service attack.

In this research, we initially provide the security community with comprehensive cyber threat mitigation taxonomy. This taxonomy, which identifies which mitigations can be automated by security controls, is the outcome of a thorough and current literature review. This implies that we're providing security professionals a concise, well-organized manual on which tactics to automate for efficient cyber-attack response.

Second, we present an automated cyber threat mitigation architecture created especially to defend against DoS attacks in the fog computing environment. By automating the protection against disruptive DoS attacks in the fog computing environment, this system provides a workable solution. In another way, it's like providing enterprises with a useful tool that automatically handles a specific kind of cyber-attack, enhancing the strength and adaptability of their defense plans to the ever-changing cyber security scene.

## 1.13 Outline

The structure of the paper follows the arrangement: Background is provided in second section however the related work of some research scholars is discussed in the third section. Fourth section includes the mapping review findings, while the methodology of our research is provided in the fifth section. The results are summarized in the sixth section. While the conclusion & future work are discussed in the seventh section.

## 2. Research Background

With the massive progress in technology, cyber security has become one of the major concerns for all organizations, governments, and other stakeholders. The most common attacks faced by organizations include phishing attacks, spam, malicious software, and malware. However, in the past few years, the extent of malware and spam attacks has increased significantly. So far, a lot of work is done in the domain of cyber-attack detection. To mention a few, the malware attacks are detected with an accuracy of 99.78% [32], DDOS attack detection rate of 99.99% [33], and spam detection with an accuracy of 97% [34]. Many mitigations are currently proposed for various attacks; however, these mitigations are tailored to a single type of attack.

The majority of the available taxonomies are for cyber threat detection. Whereas, only a few taxonomies are available for cyber threat mitigations. Currently, ATT&CK MITRE [35] is the only available taxonomy of cyber threat mitigations that also maps them to the cyber kill chain model. It is currently under development and reports a few cyber threat mitigation strategies. There is a need to build a comprehensive cyber threat mitigation taxonomy based on the existing state of the art cyber-attack solutions reported in the literature.

The cyber kill chain model consists of seven phases as shown in Table 1: reconnaissance, weaponization, delivery, exploitation, installation, command and control, and actions on the objective. In the reconnaissance phase, the attacker gathers information about the target as well as tactics for the attack [25]. In the weaponization phase, the attacker prepares their "weapons" by examining the target victim for vulnerabilities [36]. The weapons are malware payloads that are delivered in the delivery phase to the target victim's vicinity. The attackers use a variety of techniques, including USB devices, e-mail attachments, and malicious websites to spread the malware infection [37]. To acquire complete access of the victim's network or the command of the main server, attackers will exploit vulnerabilities in the exploitation phase through the delivered malware. This can be done quietly by just scanning and evaluating the target victim, but it can also lead to a direct influence on the productive of the network and systems. In the installation phase, the attacker begins to persist the infection, as well as access to the system [36]. Once the malicious malware is installed successfully in the network, it grants the attacker access to the network and the systems which ultimately leads the malware to control the network in command and control (CC) phase. In the action phase, the longer the attacker has access to the system, the larger is the impact of the attack. Additionally, ATT&CK MITRE further divides the cyber kill chain phases into 12 categories such as resource development, initial access, execution, privilege escalation, defense evasion, credential access, lateral movement, persistence, discovery, collection, exfiltration, and impact. In Table 2, cyber kill chain phases and their subdivision by ATT&CK MITRE is given.

*Table 2: Subdivision of cyber kill chain by ATT&CK MITRE*

|  | Cyber Kill Chain Phase | | | | | | |
|---|---|---|---|---|---|---|---|
|  | **Reconnaissance** | **Weaponization** | **Delivery** | **Exploitation** | **Installation** | **Command And Control** | **Actions On The Objective** |
| **ATT&CK MITRE Subdivision** | ------- | Resource Development | Initial Access | Execution | Persistence | Exfiltration | Impact |
|  |  |  |  | Privilege Escalation | Discovery |  |  |
|  |  |  |  | Defence Evasion | Collection |  |  |
|  |  |  |  | Credential Access |  |  |  |
|  |  |  |  | Lateral Movement |  |  |  |

In the resource development phase, attackers create, purchase, or compromise/steal resources that can be utilized to enable targeting. Initial access refers to a set of methods for gaining a foothold in a network that involves a range of entrance vectors. To gain a foothold, targeted spear-phishing, and exploiting vulnerabilities on public-facing web servers are the two approaches that are employed. The execution phase includes approaches that produce adversary-controlled code on a local or remote system. Malicious code execution techniques are frequently combined with techniques from other tactics to achieve broader objectives, such as network exploration or data theft. In privilege escalation, the attacker attempts to gain higher-level permissions on a system or network. Adversaries can frequently enter and explore a network with unprivileged access, however they also need elevated permissions to complete their tasks. The attacker is trying to stay undetected in defense evasion. Credential access is a set of approaches for acquiring credentials such as account names and passwords that the attacker is trying to evade. The attacker attempts to navigate through the surroundings in lateral movement. In the persistence phase, the attacker attempts to keep their foothold in the network. The discovery phase refers to the ways an attacker might use to learn more about the system and internal network. These strategies aid attackers in observing their surroundings and orient themselves before deciding how to respond. The collection phase refers to the strategies attackers may utilize to gather information as well as the sources from which the information is gathered that are relevant to achieving the adversary's goals. Exfiltration includes techniques that are used by attackers to steal data from the network. Adversaries frequently bundle data after collecting it to evade discovery while erasing it. The attacker is attempting to modify, disrupt, or destroy the data and systems.

### 1.1 DDoS Attack and Mitigation
A Distributed Denial of Service (DDoS) attack is a coordinated assault orchestrated by exploiting numerous compromised devices, often referred to as bots. These bots, infected with malicious software and under the control of the attacker, are directed to inundate a target with a barrage of data packets. These attacks can take various forms, such as overwhelming network infrastructure like servers, routers, and switches with a massive influx of traffic. Examples include TCP SYN flooding, UDP flooding, and ICMP flooding. Another form involves disguising the attack traffic to resemble legitimate data, exploiting vulnerabilities in protocols to deplete resources. For instance, attackers might target the HTTP protocol to exhaust server resources supporting web services. Distributed Reflective DoS attacks, a subset of DDoS attacks, exploit the bandwidth of victims by falsifying the victim's IP address, prompting public servers to flood the victim with legitimate responses. DNS amplification is one such attack, where small queries from compromised nodes trigger amplified responses from DNS servers directed at the victim's spoofed IP address, thereby intensifying the attack traffic [27].

Detecting and thwarting DDoS attacks at the source offers advantages in safeguarding network resources but poses challenges in distinguishing between legitimate and malicious traffic, particularly when the volume of traffic at the source appears normal before converging at the victim's end. The distributed nature of DDoS attack sources complicates accurate detection and filtering. Various defense strategies include ingress router filtering, which blocks attacking packets based on their IP addresses, and normal flow models.

However, reacting to DDoS attacks at the victim's destination is often too late, as it may have already overwhelmed processing resources and communication bandwidth. Effectively mitigating DDoS attacks necessitates a collaborative effort across the network, involving all parties beyond just the victim. Network-based mechanisms, such as router-based filtering and IP traceback mechanisms, offer more accurate responses to attacks. Mitigation strategies centered on fixed locations, whether the source or destination, may require excessive provisioning to handle unexpected traffic spikes and diverse attack types. Cloud-based mitigation presents advantages due to its abundant resources, flexible resource allocation, and high computation and storage capabilities. Cloud-based mitigation involves rerouting traffic to cloud servers for processing and filtering before reaching the destination network. Architecture proposals for cloud-based firewalling services leverage the scalability and dynamic resources of the cloud to combat DDoS attacks effectively.

Network Function Virtualization (NFV) further enhances flexibility by decoupling network function implementation from hardware, supporting agile instantiation and scaling of network functions. NFV allows for separating network functions from the underlying hardware, making it easier to adapt to changing needs. Additionally, DDoS mitigation approaches based on Fog computing offer solutions tailored to IoT environments. These approaches leverage edge computing to capture traffic profiles, aggregate data across multiple nodes, and detect DDoS attacks based on predefined thresholds. Fog computing frameworks also minimize response times by using distributed monitoring and detection mechanisms across edge and fog computing layers. Experimentation with snort rules and switches demonstrates the efficacy of these approaches in detecting and mitigating DDoS attacks.

In our approach based on Fog computing, we implement a localized, distributed, and coordinated processing system within three-level data analysis architecture. This involves conducting inline traffic filtering using field area firewalls, performing specification-based traffic analysis through virtualized network functions on local servers, and centralizing coordination via a cloud server. This centralized coordination aims to consolidate information gathered from distributed local servers, thereby enhancing detection accuracy.

3. Literature Review

4. DDOS Mitigation Framework

The stated problem that was unveiled here stems from the inadequacy of a holistic taxonomy for cyber threat mitigation, and lack of dedicated framework to identify automation-based strategies. This resource gap highlights the crucial demand in cyber security where most of global methodologies focus simply on threat detection without offering automated responses mechanism. As an answer to this weakness, the presented approach aims at offering a systematic solution that addresses these problems in totality. This methodology sets out to address different aspects of automated cyber threat mitigation by dissecting the problem into three precisely outlined phases. This organized method not only addresses the current gap in cyber defense strategies but also provides a foundation for further progress towards automated response systems.

In our methodology, we focus on designing an elaborate cyber-threat mitigation architecture aimed at effectively addressing Distributed Denial of Service (DDoS) attack. This architecture is a realization of the previous taxonomy and classification used for mitigations performed practically by an automated framework. This architecture wants to strengthen systems against DDoS attacks by automating the detection and response. Figure 1 provides a graphical demonstration of the smooth transition between processes, from taxonomy generation to mitigation classification and application of an automated framework. The integrated phases as a whole form part of an overall solution to deal with the intricate management issues that are tied up in managing cyber threats. With this methodical approach, organizations can improve their defense mechanisms and also reduce the negative impact of cyber-attacks on their work more efficiently.

### 1.2 Fog-enabled DDoS architecture

In this mitigation architecture as shown in figure 7, fog computing emerges as a powerful mechanism for both detecting and mitigating the threats posed by Distributed Denial of Service (DDoS) attacks within the network environment [27]. At its core, the Cloud Server stands as a centralized entity, boasting robust computational capabilities tailored to processing and analyzing voluminous network data.

Strategically positioned at the network's edge, Fog Nodes assume a pivotal role in the DDoS mitigation process. Functioning as intelligent gateways, these nodes harness the principles of Network Function Virtualization (NFV) to conduct comprehensive analyses on incoming traffic. This multifaceted examination encompasses a spectrum of techniques, ranging from protocol specification scrutiny to statistical analysis of DDoS features, including packet type, size, and destination target. Furthermore, deep packet inspection delves into packet header and payload content, while behavioral analysis evaluates response times, data path flows, and MAC addresses. Suspicious anomalies, such as TCP SYN flood attacks, are swiftly detected during these exhaustive examinations.

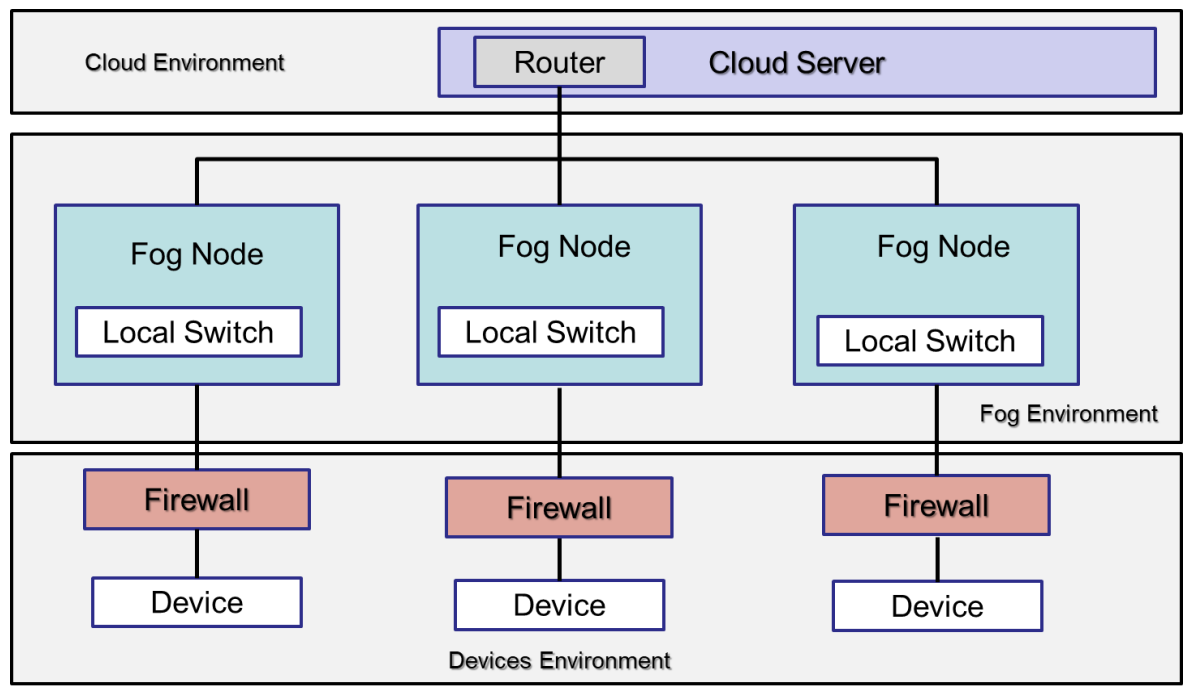

*Figure 7: DDoS mitigation architectural illustration*

To fortify the network perimeter, Firewalls are strategically deployed at the device layer. These vigilant sentinels intercept inbound traffic, subjecting it to rigorous scrutiny against predefined firewall rules. Criteria such as source/destination addresses, protocols, and port numbers are meticulously evaluated. Should any packet be identified as a known DDoS threat, immediate action ensues, with options ranging from packet dropping to logging or alert. Conversely, legitimate traffic is seamlessly routed to the appropriate fog nodes for further analysis. At the network's edge, Devices serve as the primary endpoints generating network traffic. This broad category encompasses an array of interconnected entities, including computers, servers, and Internet of Things (IoT) devices. Their active participation in the network ecosystem underscores the significance of robust security measures implemented at the device layer.

In essence, this fog computing-based DDoS mitigation architecture represents a paradigm shift in network security paradigms. By leveraging the synergistic capabilities of fog nodes and cloud resources, it not only enhances the network's resilience against DDoS onslaughts but also ensures optimized traffic management and swift mitigation responses, thereby safeguarding the integrity and availability of network services.

### 1.3 Fog-driven DDoS mitigation scheme

In this innovative fog computing-based DDoS mitigation scheme, three distinct stages are orchestrated to combat Distributed Denial of Service (DDoS) attacks effectively. Each stage employs specialized techniques and mechanisms to detect, analyze, and mitigate potential threats. The scheme operates seamlessly across the device layer, fog layer, and cloud layer, ensuring comprehensive protection of the network infrastructure [27].

#### 1.3.1 Rule-Based Packet Filtering

At the device layer, incoming traffic from devices undergoes rigorous scrutiny by local firewalls. These firewalls compare the incoming packets against predefined firewall rules, which encompass criteria such as source/destination addresses, protocols, and port numbers. If a packet matches the characteristics of a known DDoS attack, immediate action is taken, including packet dropping, logging, or alerting. Conversely, legitimate traffic that does not trigger any rule violations is seamlessly forwarded to the fog layer for further analysis.

#### 1.3.2 Fog-Based Local-Level Attack Detection

Upon receiving traffic from devices, fog nodes within the fog layer leverage Network Function Virtualization (NFV) to conduct in-depth DDoS analysis. This analysis encompasses a spectrum of techniques, including protocol specification analysis, statistical DDoS feature analysis (such as packet type, size, and destination target), and deep packet inspection of packet header and payload contents. Additionally, behavioral analysis evaluates various parameters such as response time, data path flows, and MAC addresses to detect anomalous patterns indicative of DDoS attacks, such as TCP SYN flood attacks. Suspicious traffic identified during this analysis is promptly flagged for further scrutiny.

#### 1.3.3 Consolidation And Correlation-Based Assessment

In the cloud layer, the flagged suspicious traffic undergoes consolidation and correlation-based decision-making processes. Cloud-based resources perform a comprehensive inspection, reiterating the analyses conducted at the fog layer to confirm the presence of a DDoS attack. If the DDoS attack is confirmed, the scheme applies predefined mitigation rules to take decisive action. These actions may include blocking or dropping the malicious packets, or even isolating and mitigating the affected devices. On the contrary, if the DDoS attack is not verified, regular operations recommence to guarantee seamless service delivery without interruptions.

This fog computing-based DDoS mitigation scheme harnesses the synergistic capabilities of rule-based filtering, fog-based detection, and consolidation and correlation-based decision-making to fortify the network infrastructure against DDoS attacks. Through the utilization of distributed intelligence and cloud resources, this scheme enables prompt identification and mitigation of threats, preserving the integrity and availability of network services.

### 1.4 Automated Cyber Threat Mitigation Framework

An automated cyber threat mitigation framework proposed that is proficient in mitigating DoS attack automatically is the primary goal of the methodology. This is a significant step that demonstrates the practical approach of the study in tackling cyber security concerns.

Figure 5 shows the procedural diagram, which gives a clear picture of how the automated cyber threat mitigation framework operates to mitigate DoS attacks

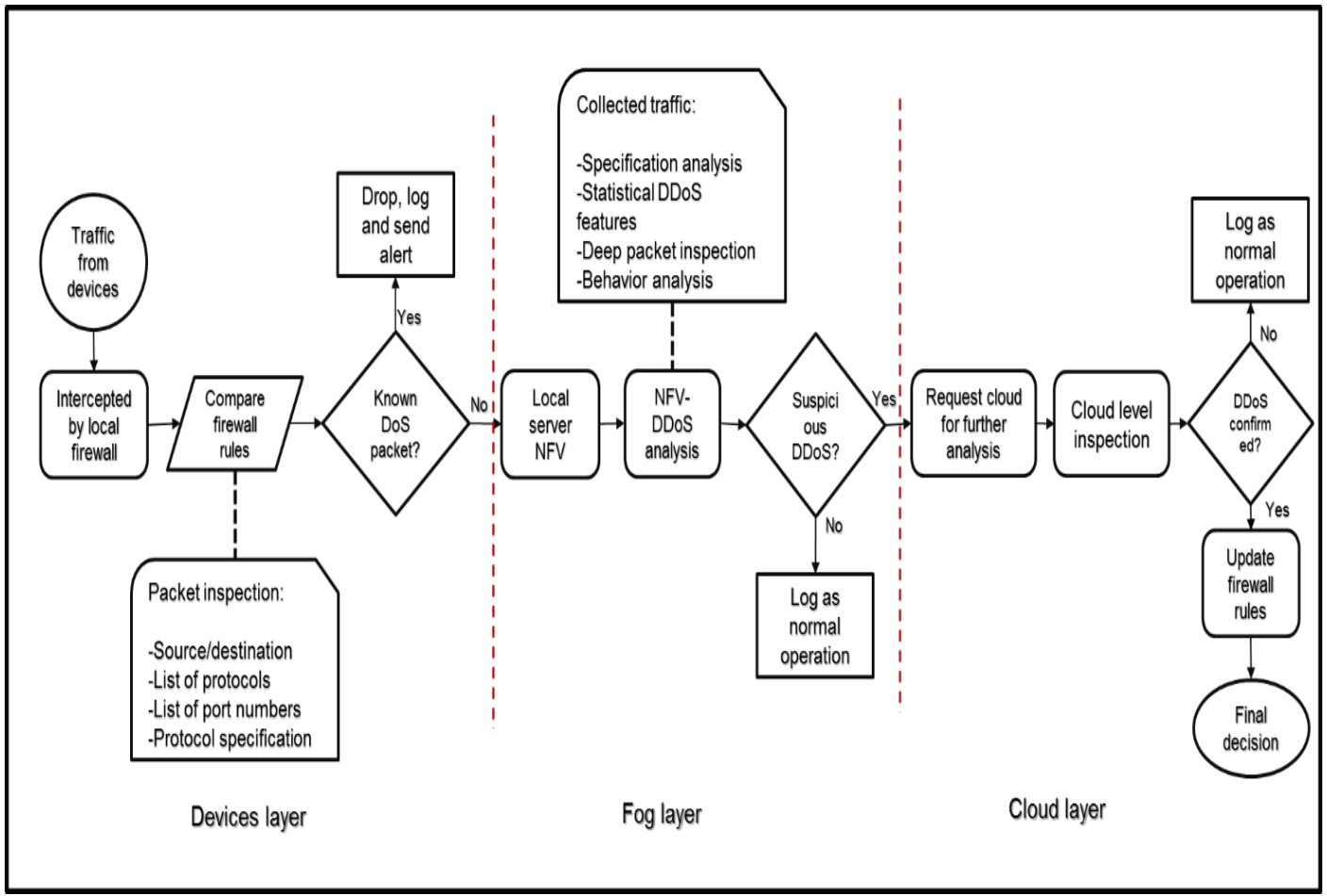

*Figure 8: Procedural diagram of proposed system*

. This diagram helps to visualize the system's operational procedure. As a concrete result of the methodological approach, this framework represents the research's dedication to filling in the gaps in cyber threat mitigation techniques and providing workable solutions for a more secure digital world. The mitigation process consists of three layers device level, fog network layer and the cloud computing layer based on providing the methodology.

### 1.4.1 Step 1: Device Layer (Initial Rule-Based Packet Firewall Inspection)

In the layer of devices, emphasis is on using intelligent appliances like mobile phones as well watches and computers to improve cyber-security. In this process, an initial scanning is done using packet firewalls based on rule applying the specific rules that have been pre-defined. Any packets that do not align with the aforementioned characteristics are immediately discarded or rejected by firewall, undertaking an essential first step defense. The active screening mechanism plays a critical role that minimizes such stages in the happenings of threats insane by focusing both to primary wave and denial-of -service. This layer is swift in recognizing and dumping suspicious packets thus helping to prevent risks associated with DoS attacks as soon as they emerge. This multi-layered approach aims to reinforce, with two operational entities collaborating and integrating their efforts aimed at strengthening the overall security structure. Together these entities fortify the efficacy of device layer aimed at defending cyber security threats.

A field device, which provides the traffic on a network, is key factors to creating communication channels over time. This traffic as it enters the device continues through an initial scanning process by rule-based packet filtering. It is a checking every packet in the systematically process that allows understanding sequence of any anomalies or potentially fraudulent activities which threatens with integrity and security for network environment. Rule-based firewall is applied to each packet and the packets are closely reviewed for any indication of DoS attacks. Through the observation of the traffic at this initial level, threats can be observed and isolated early making it easy for mitigation actions to take swiftly. Since the rule-based packet filtering mechanism plays an adversarial role, it may be operated as a preemptive countermeasure in response to mischievous activities that commonly appear within network system. By performing a systematic inspection on the incoming packets, firewalls are able to spot patterns suggesting DOS attacks such as an uncontrolled rate of traffic or abnormal packet behavior. This allows the system to implement security policies effectively so that legitimate traffic can move while putting into block or flagging off potentially bad packets. In the end, this early identification and elision process strengthens the network by making it more resilient to cyber threats protecting critical systems of data integrity in terms of sustainability.

Following the firewall in detecting and stopping suspicious Denial-of considerable traffic is bound toward a localized serving server or fog node utilizing features of Network Function Virtualization (NFV). This crucial step provides a thorough analysis of the restored traffic, especially regarding discovering hidden Distributed Denial-of-Service (DDoS) operations. The system gets the ability to explore traffic diverted near servers or fog nodes for evidence of malicious activity it left after having evaded initial security checks by firewalls. In addition, with the inclusion of fog node it adds a new layer or level which alleviates from perceived threats and is also effective in case such security emerges. The fog node further carries out an in-depth analysis of the traffic focusing on any abnormalities that could have possibly gone undetected during the original check performed by a firewall. This dual-layered approach reinforces the system's ability to detect and eliminate a wider variety of threats, thus ensuring that it is more resilient towards targeted attacks.

**1.4.2    Step 2: A Fog Network Layer-based Defense mechanism**
In this case, after the local authentication it is enabled by the traffic incorporation of Network Function Virtualization (NFV) functionality that strengthens cyber defense system with a dynamic and scalable security layer. The integration with NFV increases the system's flexibility toward new threats offering background protection against hidden Distributed Denial-of Service attacks. However, this second stage of inspection takes special importance in cases where there is an increased likelihood that attackers may use sophisticated methods to avoid being detected by firewalls. The utilization of NFV technology enables the system to dynamically detect and process incoming traffic data, thus allowing it to identify even subtle structural anomalies reflective of DDoS attacks. This responsiveness allows the cyber threat mitigation system to detect threats even beyond what has been considered normal. With the addition of an extra layer of intelligence, NFV integration strengthens and broadens protection against a cyber-attack thus achieving enhanced defense resiliency.

In the Fog Network Layer, a rigorous investigation of Distributed Denial of Service (DDoS) attacks is carried out covering aspects; so that fog node has strong resistance against such threats. This level consists of detailed DDoS analysis which is statistical characteristic-based; typically, packet type and size along with target destination are considered. Furthermore, behavioral analysis is used to analyze network traffic data looking for indication of malicious activity by taking into consideration characteristics such as source and destination Ip addresses; latencies responses times, fluid paths flows local station media address. Additionally, specification-based analysis as the methodology records protocol information being monitored to determine whether any abnormalities or deviations are observed from normal patterns of communications. Another key issue discussed in the document is deep packet inspection that studies of packets' content, including headers and payloads during inspections. This process tries to eliminate any protocol infractions, spam-like activities or viruses the result of this is that it leaves only secure content. Each of above mentioned analytical approaches serves an important function in strengthening the defenses against DDoS attacks, and preserves integrity reliability network communications.

### 1.4.2.1 Statistical DDoS Features

In this scenario, statistical analysis serves as a vital tool for scrutinizing key attributes of network traffic, allowing the system to discern between normal and potentially hazardous patterns. By focusing on packet type, size, and destination target, the system can effectively differentiate between benign traffic and potential threats. Specifically, the utilization of Distributed Denial of Service (DDoS) pattern variants is deliberately employed to train the system in identifying common attack patterns. Through this approach, the system gains valuable insights into the typical characteristics of DDoS attacks, enabling it to develop a robust understanding of the diverse range of tactics employed by malicious actors. By leveraging statistical analysis to identify patterns indicative of DDoS activity, the system enhances its ability to swiftly detect and mitigate threats, thereby bolstering the overall security posture of the network.

During the network analysis process, each packet is subjected to a rather rigorous study that helps us establish its role within the traffic in question. Such classification enables a deep the analysis of parameters in individual packets such as speed; at which can be clearly distinguished and directionality that substantially improves overview on overall network functionality. The classification of packets into various types allows the system to comprehend peculiar communication patterns appropriate for each part of communication. First of all, the packet classification is essential in determining TCP SYN flood DDoS. With the analysis of packet types, system is possible to detect specific abnormal patterns correlating with SYN flood attacks as attackers generate a lot amount of TCP SYN packets in comparison with responses that server's resources and prevent it from responding properly. Packet classification in this system provides the ability to distinguish between normal TCP handshake packets from SYN flood attack; thereby it can identify and counter such malicious attacks. The proactive approach towards intercepting the security packets facilitates in safeguarding different types of DDoS attacks thereby ensuring integrity and availability of network resources.

Yet another essential metric for the identification of network anomalies, including TCP SYN flood DDoS attacks, is packet size. The system tediously watches the packets in size that are traveling over network giving valuable information on volume and type of data being communicated between computers. The analysis of the size packets helps this system determine irregularities or unusual trends which might hint to possible differences made purposely for the purposes of fraudulent activity. In particular, regarding the TCP SYN flood DDoS attack capable of flooding a target server with too many connection requests so that it needs to reply, abnormal packet sizes can be indicators. Accordingly, attackers may tamper with packet sizes to miss utilize vulnerabilities propagated by the server's TCP/IP implementation as a resource consumption option; thru this way they would be depriving it of its usual manners. Thus, through closely monitoring packet sizes the system can easily identify deviation in size from what are normal and alerting administrators of a probable SYN flood. In addition, the system can detect other types of behavior deviations in network with use packet size analysis including anomalous instances data transfer rates or unexpected changes of values to parameters characterizing traffic activity. Through the continuous monitoring of packet sizes and comparing with pre-determined thresholds or baselines, therefore proactively detects and respond to various forms of malicious activity which consequently improves network security posture.

From the standpoint of statistical analysis, another significant issue is directed to where network traffic goes. Destination target-analysis is an integral element in the detection of TCP SYN flood DDoS attacks and other possible network threats. Upon analyzing the network traffic destinations, this system in fact gets useful information about communications trends that transpire within the said network. This analysis assists the system in identifying legitimate compared to a suspicious data flow patterns so that it can detect threats hidden behind such locations which are illicit or unpredictable, respectively. In particular, destination target analysis refers to the examination of data recipients within a network. Through the observation of common communication patters and destinations within the network, its baseline traffic behavior is established by this system. Such deviation from what has been stable over time raises suspicion and draws even more attention. Under TCP SYN flood DDoS, the attacker may try to inundate the target server with a huge count of connection requests thus jamming up all its resources. Destination-target analysis is capable of detecting such attacks by pinpointing abnormal or unauthorized destinations that are subjected to an alarmingly high level of traffic. The deviations from normal patterns of traffic behavior signify hazards associated with malicious activity that should require immediate remedy. In general, a result of destination target analysis makes it possible to improve the performance in detecting and neutralizing network threats as anomalous traffic patterns can be identified with potential security risks detected. The system continuously monitors destination targets with the comparison of firewall rules as it involves pro-active protection against TCP SYN flood DDoS attacks.

Within the network and justified with comprehensive analysis of packet size, type and destination targets this strategy proves an effective tool for resilient threat detection mechanisms. Through analysis of the identified attributes, then system conducts a statistical noise around data traffic amounts for several transport layers to evaluate patterns in network activities. Such a holistic approach allows the system to take informed decisions regarding the integrity of network activities. A useful pattern of normal behavior is that many attacks have similar characteristics from packet to packet, and by correlating features with these characteristic the system succeeds using them as discriminants. For instance, it is capable of pinpointing early indicators or the beginning battle from an emerging threat like a malignant injection TCP SYN Flood attack patterns into their network. This in-depth analysis empowers the system to recognize abnormal behaviors and security threats more efficiently. Through forward-looking analysis, network traffic monitoring and scrutiny of packet attributes it can detect threats in real time while protecting the network from a broad set of cyber-attacks. It is mandatory because of dynamic and changing nature since threat- landscapes the network infrastructure has to be kept intact with min neglecting any provision.

Under special conditions, namely when anomalies indicative of a TCP SYN Flood attack are found in the traffic it responds immediately by issuing alerts and marking such suspicious package. This manner is proactive, which corresponds to traditional cyber security practices that emphasize speedy threat detection and elimination. After detecting the suspicious traffic attacking flow, an in-depth collection of packets takes place to ascertain source and destination IP addresses being exchanged without their payload content as well as protocols together with port numbers currently used. Employing the concept of virtual network environment, this system is capable to lead one through statistical analysis techniques which uncover risks. It uses specific detection methods that effectively detect threats such as TCP SYN Flood attacks providing much-required strong cyber security in virtualized network architecture.

### 1.4.2.2 Specification Analysis

Specification analysis, encompassed within the layered cyber security strategy as applied in the fog network layer relies solely on continuous monitoring and evaluation of protocol provision to ensure that communications take place not only by adhering set standards. This procedure implies defining anticipated behaviors of attended network traffic through a wide range of rules and criteria. In this way, the system acquires an opportunity of uncovering each deviation from normalcy which would reveal a security breach or illegal activity. In addition, specification analysis is an essential aspect in the preservation of network security and integrity. The system should receive traffic on the network though this is not biologically one of its developing tools; however, as soon as detecting deviances in that type of activity lead it to flag them within grave suspicion and implemented proactive countermeasures before they will get worse. This proactive methodology serves as the defense against leaked sensitive data, prohibits unauthorized account proceedings and assures that your network infrastructure's reliability level is maintained; therefore its functioning still remains sufficient. Notably, the specification analysis further provides uninterrupted monitoring of the network to ensure that its resistance towards evolving cyber threats is monitored and any abnormalities from compliance are also continuously highlighted.

Essentially, the specification-based analysis is qualitative in determining to what extent network traffic satisfies predefined parameters. Actively the protocol compliant and finds any patterns that are associated with established cyber security threats. The direction is to identify a way of anticipating aberrations that could imply possible security danger. Its analysis of each device includes a network transmission in which frequency variables such as connection numbers are addressed and the packets generated from them, With this, it finds that an immense number of connections appear to be suggestive for a distributed denial of service (DDoS) type attack; specifically a TCP SYN flood. Moreover, the network does not create traffic for one device and follows strict standards of separating malicious packets from those that adhere to protocol characteristics. What is characterized as a case of TCP SYN Flood attack occurs upon the observation that, more linkages than commons are made by one terminal on any given network. A flag is set based on an event where the behavior deviates from normal and marks all network traffic attached to that device as questionable. it then uses this anomaly detection mechanism to immediately forward the suspicious packets (that are thus identified) towards a centralized cloud server in order for more detailed investigations. The cloud's resources and expertise are used by the system to evaluate these identified packets; hence, this collaborative method is utilized.

Finally, suspicious attack traffic is deeply analyzed by the cloud server that may necessitate use of more complex algorithms and resources. This collaborative approach allows for further analysis of the detected packets, an assessment to determine their nature and function as well as the ability network effect. The outcomes of this investigation facilitate better threat intelligence and levels overall cyber security by establishing the value amass specification-based analysis in identifying trying to isolate, detecting probable security threats inside network.

### 1.4.2.3 Behavioral Analysis

Behavioral analysis, considered a sophisticated approach to threat detection inside network traffic, is essential to enhancing cyber security defenses. This approach analyzes important characteristics such as source and destination IP addresses, response times, data channel flows, and MAC addresses, going beyond standard detection methods. By combining these characteristics, we hope to identify minute deviations from normal network behavior that could indicate possible malicious activity. This research is useful in the sense that it can help to find risk factors that could occur and remain hidden from traditional cyber security measures.

Namely, the behavioral features that are added in this system enable a better detection of slight deviations in network activities which are aimed at capturing the advanced variants of cyber force that can fly past the current operational scans. This method is appropriate for the needs because it is pragmatic; behavioral analysis is a major part of identification and response to new threats in the field of network security. Furthermore, behavior analysis also utilizes multiple scenarios through the analysis of network features for varying response behaviors. Applying parameters such as the response time, data channel flows and unique MAC addresses to this approach allows for a more in-depth analysis of the stereotypical network behavior in its numerous directions. By allowing the system to identify deviations from normal pattern of behavior, this in-depth evaluation helps to improve the system's threat detection capability.

In the field of behavioral analytics, the introduction of specific elements like response time, data channel flows, and unique MACs is a purposeful measure that was used so that to understand how such potential threats from cyberspace might be detected by this system.

Response time leaves an important role to detecting TCP SYN Flood DDoS attacks, which can be considered one of the crucial signals from such factor as anomalies in network. During normal network operation if a device sends data transmission, it looks for the timely receipt of response at destination side. But when in a context of TCP SYN Flood attack, where an attacker sends many requests that lead to IP address overflows the network can be overwhelmed. Therefore, the average response time for valid data transmissions may increase drastically caused by connect request deluge congestion; such abnormal increases can be detected by closely monitoring response times of the network administrators. The abnormally high response time shows that the bandwidth is overused or there are someone else's attempts to load down system. Recognition of abnormal responses to sequences and variations constitutes detection indicators, which aid administrator in recognizing the presence TCP SYN Flood attack hence taking appropriate countermeasures. Through the measures to contend attack effect, by constraining bad traffic or limiting connection demands administrator can return network's performance and dependability. Therefore, response time analysis serves as a useful tool to gain more understanding of behaviors within the network and allows for early detection of TCP SYN Flood DDoS attacks effectively thus making networks stronger in their ability to withstand such threats.

Flows in the data path simply means that each of this unique routes or pathways which are considered as flows on which they travel during their tripping across to contains some discernible flow These paths are individual communication lines between network elements. When such flows are analyzed, the network administrators gain valuable data about interaction volumes and complexity in a given network. So, as it applies to TCP SYN Flood DDoS attacks detection, monitoring flows in the data path is very important. As part of a TCP SYN Flood malicious user launches continuous flow to floods the network with generated connection requests in an attempt to exhaust all available resources and interfere obstructing normal traffic. Therefore, the number of flows in data path can have a spike that reflects activity increase due to floods connection requests. Through that is by carefully observing the flow patterns and recording any changes from expected levels, administrators can note signs of suspicious behavior showing evidence of a TCP SYN Flood attack. The influx of flows, almost exclusively with respect to connection initiation increases are among the first things that hint about an imminent breach. Timely detection enables administrators to implement security measures such as the filtering out of malicious traffic, or rate limiting connection requests which restricts collateral damage and protect network integrity. In general, flow analysis is essential in terms of identifying and counteracting TCP SYN Flood DDoS attacks for the purpose of preserving network stability and robustness.

MAC authentication addressing implies the analysis of distinctive identifiers that have been assigned to network equipment involved in communication. The identifiers are called MAC addresses and they play an important role in tracing activity within the network as well as for identifying a particular device on that network. Detecting TCP SYN Flood attacks makes the usage of MAC address following necessary. In the process of such strikes, malicious actors usually try to inundate with a high level of anomalous connection requests sent through network. Such malicious behavior can be detected by the network administrators when they analyze MAC address activity close to monitor irregularities. For instance, rapid increase in the amount of new MAC addresses observed on a network or strange behavior patterns in terms of MAC address activity may point at conductive operation. Through careful tracking of MAC address behavior, and detection suspicious patterns in it admins have an opportunity to take preventative measures against the consequences of TCP SYN Flood Attacks. This could be achieved by ways of installing systems for filtering out harmful traffic or capping the rate at which connection requests are made. As part of its DDoS protection defense strategy, administrators using MAC address analysis reinforce the security provisions on their network and make resistance to a comprehensive attack more effective. The proposed behavioral analysis strategy seeks to improve the efficacy of cyber security defenses, using information that is obtained from response- time share studies; data flow diagrams and MAC address examination. The pattern identification, along with the appearance of anomalies in network behavior could provide administrators an opportunity to introduce preemptive measures that would meticulously try and prevent whatever threats for escalating. Such proactive actions may include introduction of access controls, network settings tweaks or installation systems with other security protocols. Combining behavioral thinking with forwarding action focuses, enterprises can enhance the cyber security posture and control risks caused by rising threats.

**1.4.2.4 Deep Packet Inspection**
Deep packet inspection (DPI), also known as the scanning of each and every passing network traffic data, is one intrinsic element which throws light upon a large number of important information related to monitoring and attacking protocol. In the process of analysis, this particular one deeply into payload and packet header is conducted. By packet content analysis, the system is able to block spam as well as malware cases and protocol violation instances. Deep packet inspection offers in-depth look to reveal the characteristics used and known that allows it to identify all cyber security threats as a whole. The process is essential to the cyber security strategy as it provides an exhaustive understanding of what data transmitted, and enables powerful application for threat detection methods.

To detect TCP SYN Flood DDoS attacks, deep packet inspection or DPI serves a significant role in analyzing the data contained in both the header and payload. Here's how DPI aids in this detection process: Packet headers hold considerable information on the packet's origin, recipient for instance when and domain some of them important elements it is to be noted. In deep packet inspection the header of inbound packets are inspected to determine if any anomalies or suspect attributes present within. Thus, DPI is capable of spotting more irregularities with respect to TCP SYN Flood attacks by detecting abnormally high volume of SYNs not followed up by ACKs or improper use of flags in TCP headers. DPI can help in identifying packet headers suspicious characteristics characteristic of SYN Flood attacks, which will allow timely detection and response. Among other packet header analyses, DPI takes a step further into the content of payload to yield even more information. The payload is the data which needs to be transmitted; it maybe commands messages or any type content of communication. In deep packet inspection the payload content is examined for any indication of malevolent activities or patterns that are affiliated with SYN Flood attacks. For instance, DPI can notice abnormally big payloads that include repetitive or nonsense data towards an attempt of overwhelming the target server and network. Due to DPI's ability of perusing payload content, SYN Flood attacks can be better detected and proactive mitigations measures could have taken. In all, deep packet inspection is the centerpiece in fighting TCP SYN Flood DDoS attacks by critically examining both headers and contents of packets. Thanks to the DPI technology, organizations can easily detect and counter SYN Flood attacks therefore secure their networks from total disruption of services.

**1.4.3   Step 3: Cloud-level Inspection**
In the process of mitigating cyber threats, cloud inspection serves as a critical pillar that imparts another level of review to ensure reliability and efficiency in threat detection. Although the fog nodes perform some initial analysis of incoming traffic, it is in the cloud server that this suspicious activity is thoroughly examined. This dual approach guarantees deep network traffic analysis from multiple perspectives. Cloud inspection validates findings from the fog sensors and confirms DDoS behavior, collecting data from all fog elements to provide behavior patterns characterizing the entire network infrastructure. This in-depth analysis fosters a detailed understanding of network activity, aiding in the identification of subtle signs of DDoS attacks that may have evaded detection at the fog node level. By correlating data from various sources, cloud inspection reduces false positives, ensuring that only genuine threats are addressed. This cooperative approach improves the overall proficiency and precision in DDoS detection, allowing for a comprehensive understanding of network dynamics and facilitating timely response to potential threats.

In our proposed methodology for network security, we focus on mitigating potential threats, particularly Distributed Denial-of-Service (DDoS) attacks, by integrating cloud inspection and proactive mitigation strategies. Cloud inspection serves as a critical step in validating and confirming suspicious traffic patterns identified by fog nodes, helping to detect any signs of malicious behavior within the network. Once confirmed, the cloud server initiates proactive mitigation measures, such as dynamically adjusting firewall policies based on observed packet patterns from DDoS attacks, to block or drop suspicious packets in real-time. Additionally, we incorporate a set of mitigation rules within the network infrastructure to enhance its resilience against cyber threats. These rules include detecting and mitigating common attack vectors, such as TCP SYN floods and abnormal response times, through targeted actions aimed at neutralizing potential threats before they can disrupt network operations. Some of the rules for mitigation include:

Rule 1: TCP SYN Flood:

drop tcp any any -> any any (flags: S; msg:"TCP SYN flood detected!"; sid:100001;)

This rule drops TCP packets with the SYN flag set, indicating a potential SYN flood attack, and generates an alert with the message "TCP SYN flood detected!".

Rule 2: Abnormal Response Time:

network activity.alert any any -> any any (threshold: type both, track by_dst, count 50, seconds 1; msg:"Abnormal response time detected!"; sid:100002;)

This rule triggers an alert when an abnormal response time is detected, based on a threshold of 50 occurrences within 1 second, indicating possible performance degradation or suspicious

Rule 3: Excessive TCP SYN Packets:

drop tcp any any -> any any (flags: S; count 10; seconds: 1; threshold: type both, track by_dst, count 50, seconds 1; msg:"TCP SYN flood detected!"; sid:100008;)

This rule drops TCP packets with the SYN flag set and a count of 10 occurrences within 1 second, indicative of an excessive number of SYN packets and potentially a SYN flood attack.

Rule 4: TCP Packets with High Number of Half-Open Connections:

drop tcp any any -> any any (flags: SA; count 5; detection_filter: track by_dst; msg:"Drop TCP packets with a high number of half-open connections"; sid:100003;)

This rule drops TCP packets with the SYN-ACK flag set and a count of 5 occurrences, indicating a high number of half-open connections, which is characteristic of SYN flood attacks.

By combining cloud inspection, proactive mitigation measures, and the integration of cyber-attack detection rules, our methodology provides a robust framework for safeguarding network infrastructure against a wide range of cyber threats, including DDoS attacks.

Cloud inspection plays a crucial role in affirming and responding to potential Distributed Denial-of-Service (DDoS) attacks within the cyber threat mitigation process. This involves the identification of suspicious traffic patterns by fog nodes, followed by exhaustive analysis using the cloud server to validate and confirm any detected DDoS behavior. Through the collection of information from various network sources, the cloud inspection process provides a comprehensive understanding of traffic patterns and uncovers any anomalies indicative of malicious intent. Upon confirmation of DDoS behavior, proactive mitigation measures are enacted by the cloud server, which may include adaptive adjustments to firewall policies based on observed packet behaviors. These measures effectively block or drop suspicious packets, preventing potential disruptions to network services. Furthermore, cloud inspection ensures the continuity of legitimate network activities by accurately identifying benign traffic and allowing normal communication flow. This self-defense mechanism preserves network integrity and eliminates hostile forces, thereby enhancing overall security posture. Through collaborative analysis and preventive countermeasures, cloud inspection strengthens the network's resilience against DDoS attacks, reducing the likelihood of incidents such as TCP SYN Floods.

### 1.5 Threat model

A threat model refers to a structured method that is used in the identification, assessment and mitigation of threats as well as security weaknesses within interrelated system environment [88]. It has several components such as the assets, rivalries adversaries threats/attacks countermeasures and it takes a systematic approach to analyze how secure is that system [89].

#### 1.5.1 Assets

These constitute the valuable parts or elements that consequently need security within the system [90]. At the device layer, assets include devices both of hard and software compositions as well as infrastructure and network connections. Within the fog node analysis, assets are comprised of fog Nodes, NFV infrastructure and firewall rules. In the case of cloud analysis and mitigation, assets are considered such as a cloud server that is aimed at being analyzed or bypassed through their suspension; suspicious traffic data which can be evaluated using various means including network flow records for providing real-time visibility onto any unusual communication sessions.

#### 1.5.2 Adversaries

Here we mention possible threat vectors or potential adversaries who would like to take the advantage from flaws in the system [91]. Such adversaries can concern malicious users or sophisticated attackers.

#### 1.5.3 Threats/Attacks

These are the special risks or attempts that such adversaries will employ to penetrate and compromise this system. These are unauthorized access, Distributed Denial of Service attacks or confirmation failures.

#### 1.5.4 Countermeasures

These are defensive countermeasures or the controls to neutralize effects of identified threats. The countermeasures could be firewall rules, multi-analysis confirmation, traffic monitoring and regular infrastructure maintenance.

As shown in the diagram, figure 9, threat model consists of three layers that address individual elements primarily focused on the system itself and their appropriate security considerations.

#### 1.5.5 Device Layer

Device layer includes the building blocks of the system such as various devices, infrastructure elements and network connections. Devices belonging to this layer include the devices themselves, any underlying infrastructure that supports their operation and connections between such devices through a network. Users that pose malice intent towards this will target these attackers. Such threats within this layer may take form as similar to illicit access bids or attacks upon weaknesses of devices infrastructure. The countermeasures to combat these threats usually include the implementation of complex firewall rules, traffic monitoring systems as well as regular updates in infrastructure ensuring known vulnerabilities gets addressed and system integrity is upheld.

### 1.5.6 Fog Node Analysis

The node analysis layer of the fog concentrates on security issues inherent in fog nodes and NFV infrastructure associated with them. Assets in this tier encompass fog nodes, NFV infrastructure that supports their functionality and firewall rules which govern access control as well regulation flow of traffic among these devices. Enemies focusing on this level could be attackers disrupting fog node functionality to impede the operation of downstream components or compromising NFV infrastructure and using it as a staging ground for attacks against lower network planes. Attacks such as DDoS attacks targeting fog nodes or compromising the firewall are mere threats and try to bypass rules so they could get unauthorized extracts. Such countermeasures mostly employ multi-analysis confirmation methods to confirm the legitimacy of incoming traffic, and thereby ascertain that fog node operations are secure.

### 1.5.7 Cloud Analysis & Mitigation

The layer for the cloud analysis and mitigation concerns security aspects in relation to both, on one hand, suspicious traffic data while also managing this key aspect of cracked server. The assets under this layer comprise of the cloud servers, data sets which consist included suspicious traffic details as wells Snort firewall rules utilized for intrusion detection and prevention. The enemies of this layer may include advanced attackers who want to exploit weaknesses in cloud infrastructure or bypass detection mechanisms for performing sophisticated attacks. Threats/attacks may now be confirmations failures or attacks designed to bypass mitigation measures in an attempt for launching successful ones. The countermeasures in this layer frequently include upgrading of the updates to snort rules and infrastructure keeping abreast with current threats ensuring a stabilized security posture within cloud environment.

More broadly, the threat model offers a systematic approach for identifying, evaluating and managing potential security risks associated with various layers of the system. Through systematic asset, adversary, threat and countermeasure analysis for each layer internal organizations could know their security position as well plan how to protect themselves from potential threats or attacks.

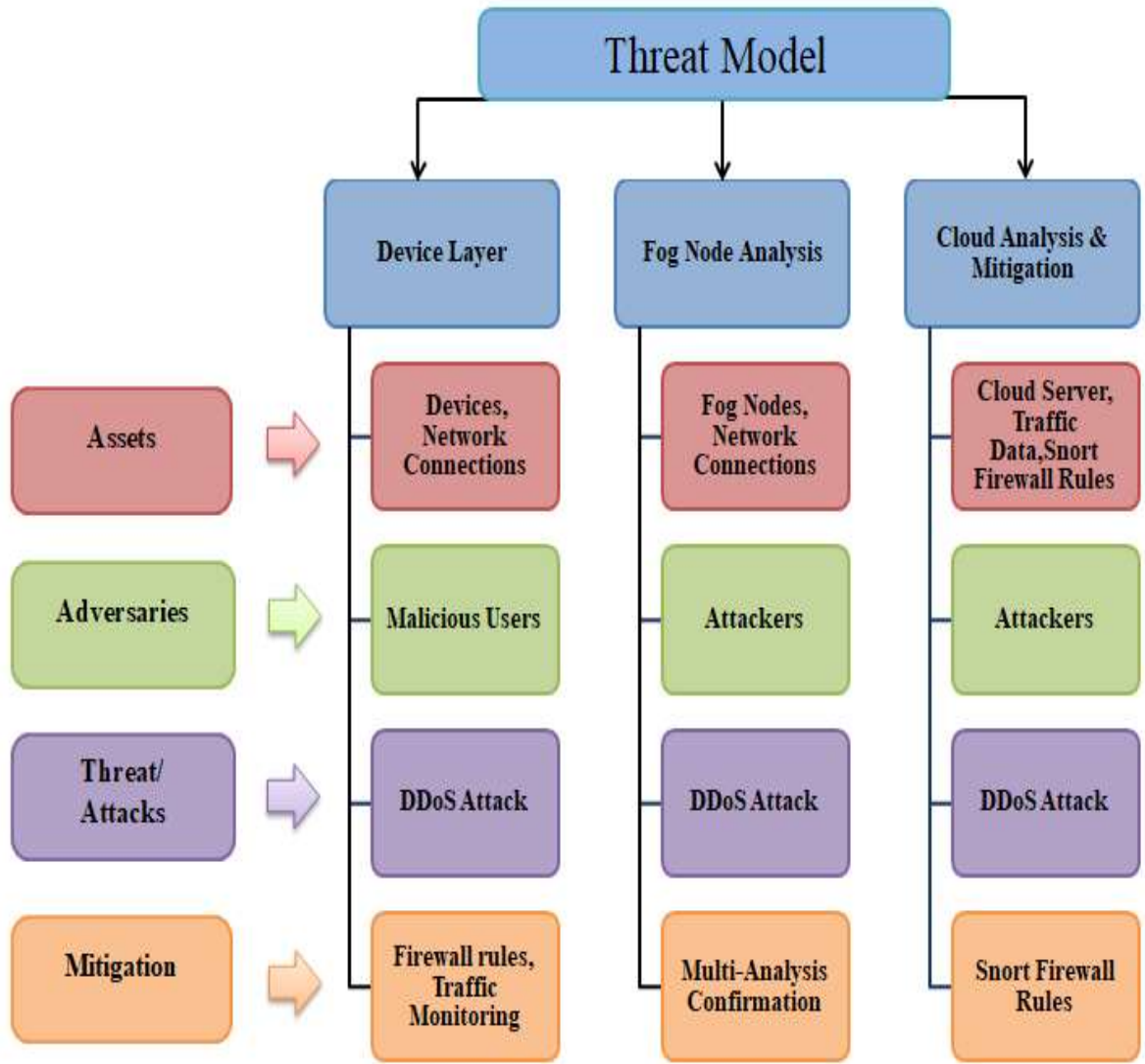

*Figure 9: Threat model of the proposed approach*

## 5. Experiment Implementation

In the section of result analysis we study a detailed assessment of the proposed framework in order to verify its effectiveness and performance used for combating cyber threats. Assessment is comprised of detection and mitigation accuracy, as well as efficiency in term sof scalability and applicability to different attack visionaries. Therefore, in a systematic manner we aim at providing an overall performance evaluation of the framework's functionalities and validity to be deployed into practice. In addition, the assessment is a comparative analysis contrasting the proposed framework with other currently prevailing cyber threat mitigation approaches to provide benchmarks of measurement against established methodologies. This comparative assessment provides useful findings relating to the strengths and weaknesses of the framework, in terms of what needs improvement or innovation.

In addition, we investigate the effectiveness of the proposed approach using different deployment scenarios and network configurations to check influence on implementation efficiency. In an effort to test the frameworks robustness and resilience in dynamic environments, we intend on simulating different cyber threat scenarios and network conditions. In general, the analysis section results play an essential role in our study emphasizing empirically based evidence and knowledge of how effective is presented a framework. By in-depth assessment and analysis, we strive to offer recommendations that are impactful on the strategies used for mitigating cyber threats as well increase the security level of network protection from fresh hazards.

### 1.6 Experimental Setup

The approach is carried out on a personal computer that has Windows 10 pro, a 64-bit operating system, an AMD A6-6310 APU with AMD Radeon R4 Graphics 1.80 GHz processor, 8.0 GB of RAM, and an x64-based processor. The Python programming language is employed in the implementation of this methodology. Also, Snort 3.x 3.1.36.0 [92] (open source intrusion detection and prevention system IDS/IPS), will be used that includes a series of rules required to monitor the malicious network activity.

### 1.7 Initial Rule-Based Attack Filtering

A network scenario consisting of ten devices and a firewall system is modeled in this simulation. The imposed rules control the movement of network traffic in that it determines source and destination IP's, protocols and ports. Through the generation of attack traffic, site is created to simulate network activity between these devices in introducing a possibility that DoS attack may arise. Figure 10 shows the traffic simulation of the devices.

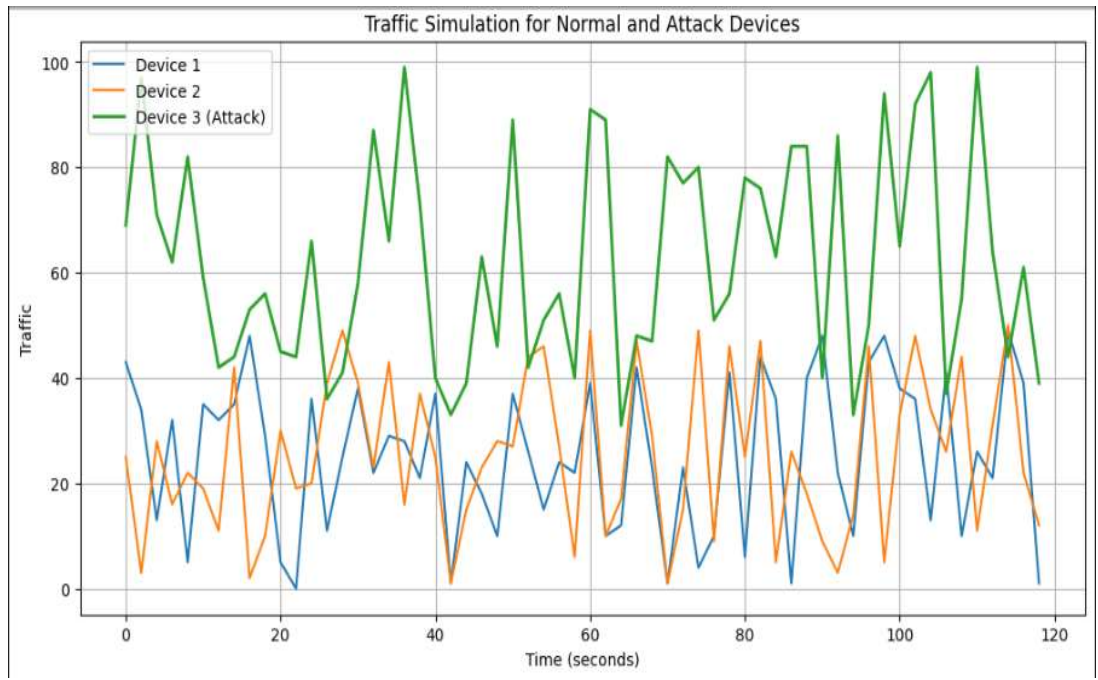

*Figure 10: Devices traffic simulation*

The introduced methods detect and respond to the simulated traffic, aiding in identifying as well as managing potential DOS packets priori through specific requirements. After the simulation, detailed statistics are given on detected DoS packets; successfully forwarded to destination and dropped a number of received or sent alerts. The packet delivery ratio is measured which specifies the ratio of transmitted packets that were successfully delivered out of all simulated.

Moreover, visuals are included to help understanding. A composite bar-and-line graph shows data (counts) for individual packet categories, which include detected DoS packets, forwarded packets dropped and alerts. Moreover, packet delivery ratio graph shows the amount of delivered packets in each device from simulated network. The presented results reveal that from 10,000 packets, 9882 were forwarded, and 118 were found as DoS attacks whereas 85 of them got dropped while alarms for other 33 have raised. The packet delivery ratio of 98.82% attests to the effectiveness of network in delivering most simulated packets efficiently. The statistics of packets and also packet delivery ratio are shown in the graphs below (figure 11 and 12).

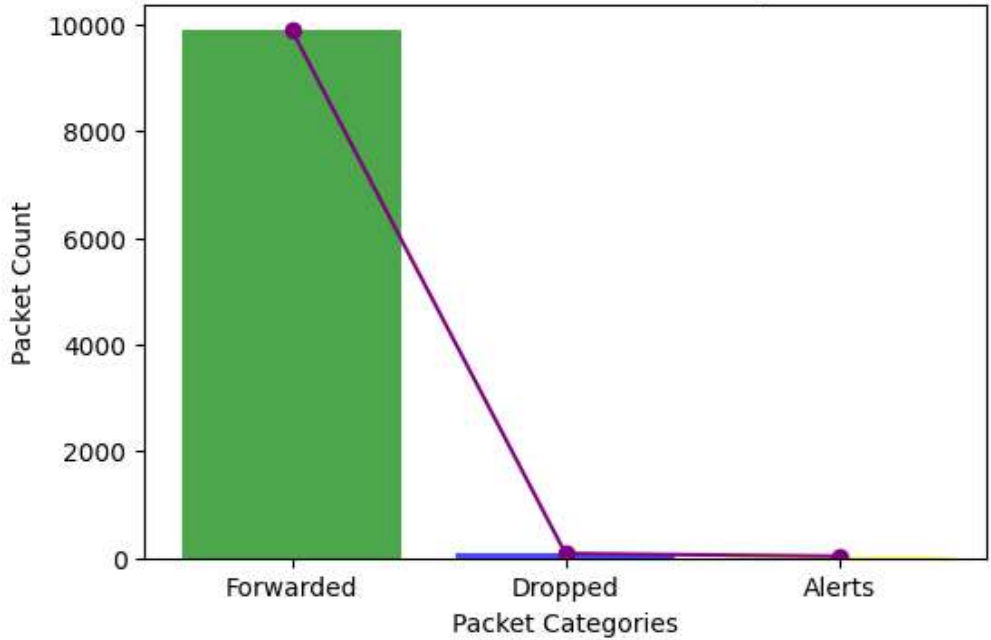

*Figure 11: Packet statistics*

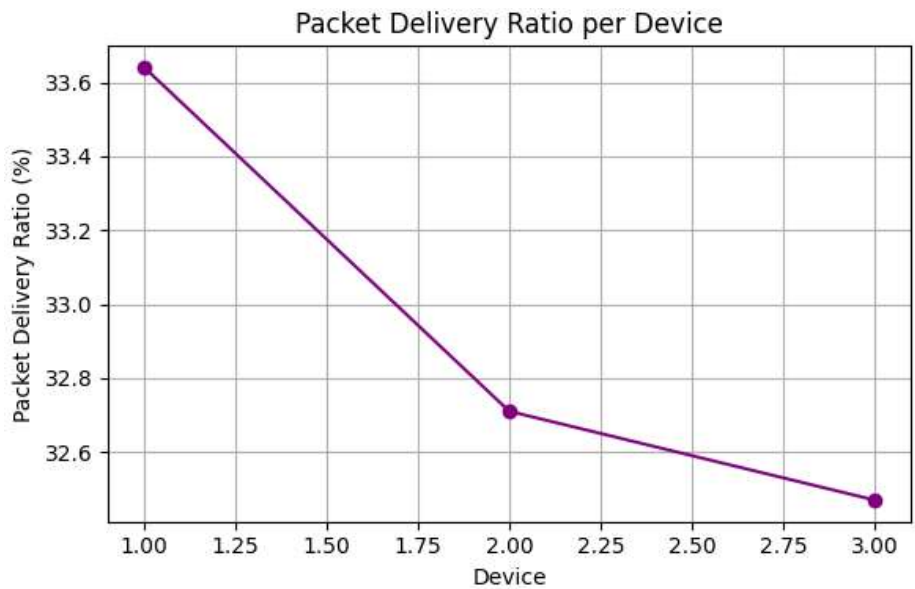

*Figure 12: Packet delivery ratio per device*

**1.8 DDoS Attack Detection in the Fog Nodes**
The implemented operations together examine network traffic within fog nodes to detect Distributed Denial of Service (DDoS), which is mainly emphasized on TCP SYN flood attacks. One part is devoted to the creation and analysis of statistical network traffic for individual devices in the fog layer. This simulation includes normal and malicious packets protecting TCP SYN flood attacks. The results are numbers of suspect devices, details on regular traffic and a summed up list of suspicious packets.

The second aspect is to implement a behavioral simulation of the networks for all devices based on response times, flow in data paths and MAC addresses. This simulates uses cases of TCP SYN packets to model an attack situation. The results inform about response times flows, MAC addresses SYN counts and complete list of suspicious packets. Similarly, another aspect is generating network behavior for each device given that reaction times and data path flows along with MAC addresses. For this simulation, the instances of TCP SYN packets are added as representative attacks. The output includes information about response times, flows in the network path, MAC addresses SYN counts and a complete list of suspicious packets. Additionally, traffic analysis is conducted on the basis of specifications to determine suspect devices and classify some subset as non-suspect. The outcomes consist of data on potentially dangerous devices, device statistics non-suspicious traffic and the list of identified suspicious packets. The strategy of DPI is being used in order to inspect network packets, particularly aimed at TCP SYN flood attacks. This operation records overhead accompanying the inspection process and discerns packets for further scrutiny that are suspicious. A broad approach incorporates sniffing and inspection of network traffic. This involves the transmission of network traffic and application DPI on each packet. The purpose of the operation is to measure DPI overhead and detect dubious traffic, outputting detailed findings for further study.

Delving deeper into the comparison of detection times, there are considerable enhancements in effectiveness by using the proposed network traffic analysis method. It does so much better than the times seen in existing literature shown to be on average 0.246s as depicted in figure13, achieving a detection speed of 0.159s. This time down reduction reveals the increased responsiveness and agility of our method. It allows more rapid identification of risk zones in the fog computing environment helping to ensure timely threat detection and handling. The improved efficiency strengthens the cyber security aspect for fog computing networks, reducing risks, and expanding detection limits.

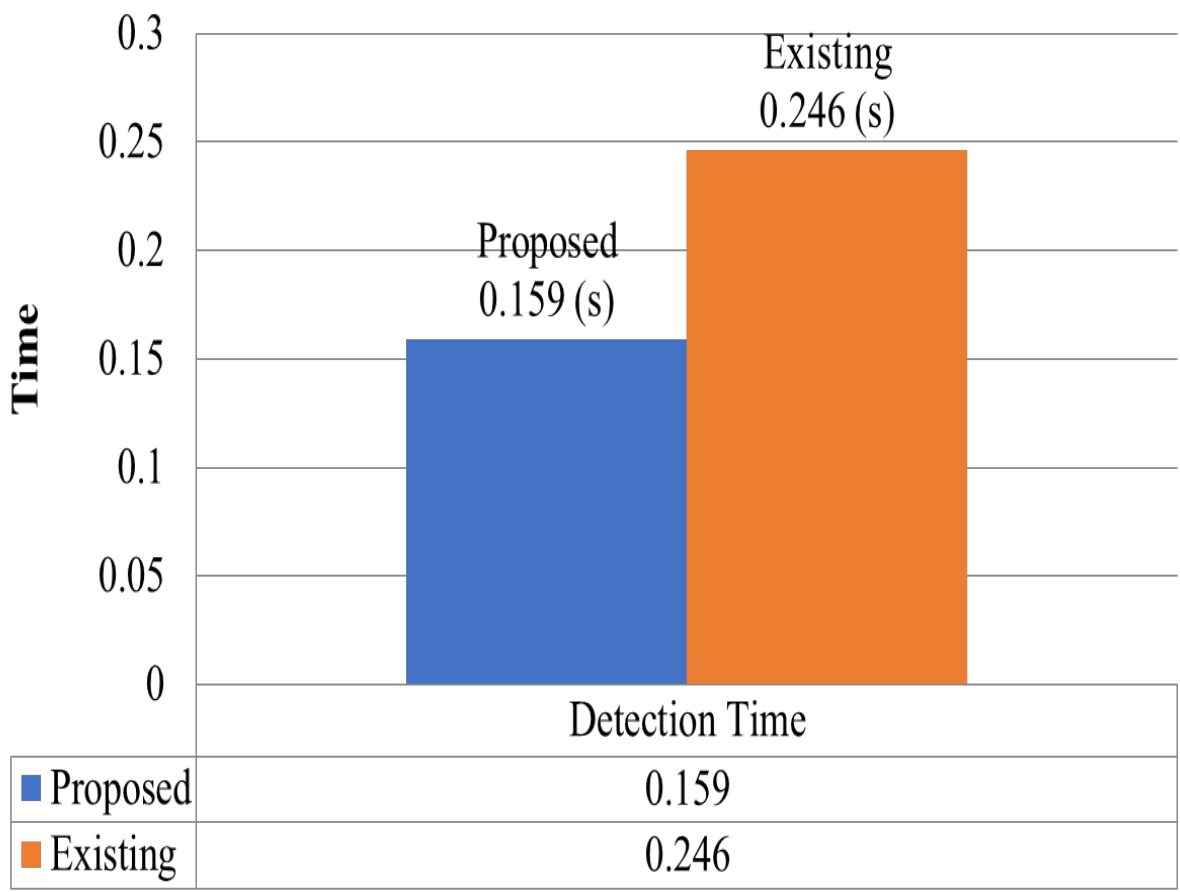

*Figure 13: Detection time comparison*

It is to be clarified further that the better efficiency shown by requirement method is mind-blowing in its very high detection rate 99.86% assignable from figure 14. This is contrary to the average detection rates of about 99.56% [27] previously reported in literature. This significantly higher detection rate has much to do with an increased degree of precision and reliability for the network traffic analysis method we are proposing. Having a detection rate that is almost perfect means only potential threats within fog computing networks are detected and marked with higher accuracy.

The visualizations presented in figures 13 and 14 provide strong support to the proposed approach's strength and effectiveness. These numbers not only corroborate the superiority of the detection rate attained but also showcase such a general effect on elevating cyber defense accurately in fog computing environments. With improvements in accuracy and reliability, the method provides strong protection against potential hazards to fog computing network.

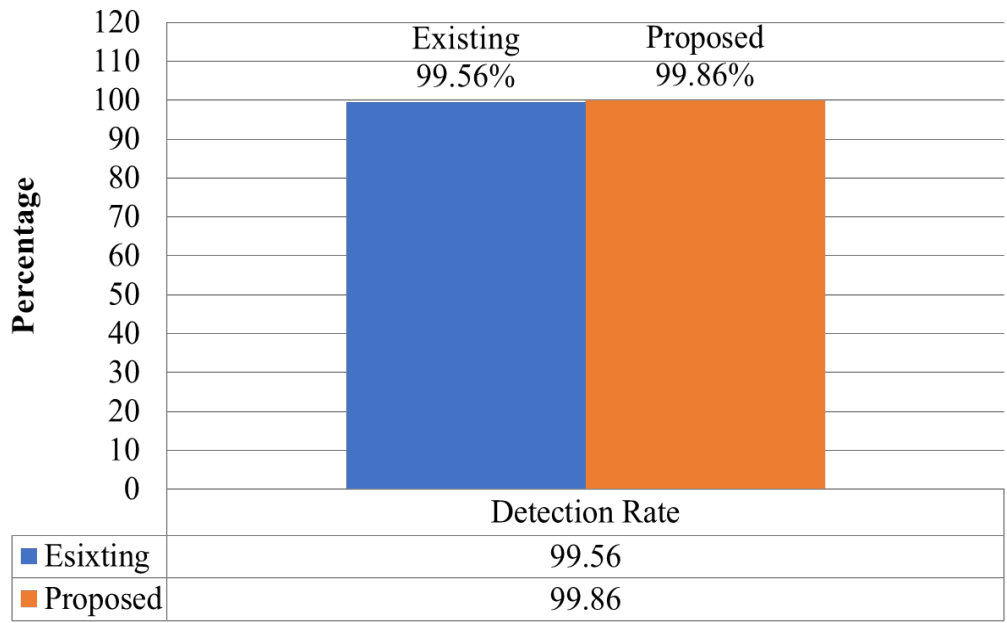

*Figure 14: Detection rate comparison*

**1.9 Advanced Cloud Investigation For DDoS Mitigation**

The integrative function dictates cloud-level monitoring to verify and neutralize DDoS attacks, specifically TCP SYN flood incidents through Snort firewall directives. From the behavior-based, specification-based, statistical based and DPI evaluations all merged insights lead to a collection of marked devices and packets. The key operation is confirming DDoS patterns based on various benchmarks, including unusual behavior features, specification-based models sets of rules corresponding to statistical anomalies and specific content filters detected through the deep packet inspection. The verified DDoS-implicated devices and packets are then subjected to more mitigation protocols. This mitigation means blocking verified DDoS implicated devices within a network graph and sending Snort-mitigating directives to confirmed flagged packets. The overall aim is to skillfully and promptly detect address any prevalent DDoS threats in the cloud server network.

The implementation includes a section devoted to Snort rules adapted for each analytical category, which are behavior analysis, specification analysis, statistical DDoS ones and deep packet inspection. These commands provide standards for measuring and preventing DDoS attacks. Secondly, the implementation involves functions for processing both lists and dictionaries of packets to be checked against Snort rules. The mitigation process captures and documents the time involved as a result of mitigating, besides representing the percentage of packets that were eliminated out to package flagged number. However, the described metrics altogether form its assessment of how efficient and effective the cloud-level scrutiny and mitigation process is.

The outcomes of the process cloud-level scrutiny and mitigation demonstrate its amazing effectiveness in protecting against possible DDoS attacks. With the quick mitigation period of 0.427 second, the system is able to react immediately in response emerging threats from DDoS attacks only minimally impacting on cloud server infrastructure environment. Moreover, high mitigation level of the snort indicated at 96.32% in figure 15 shows that most packets flagged by predefined rules were found not to be part of network traffic and hence blocked.

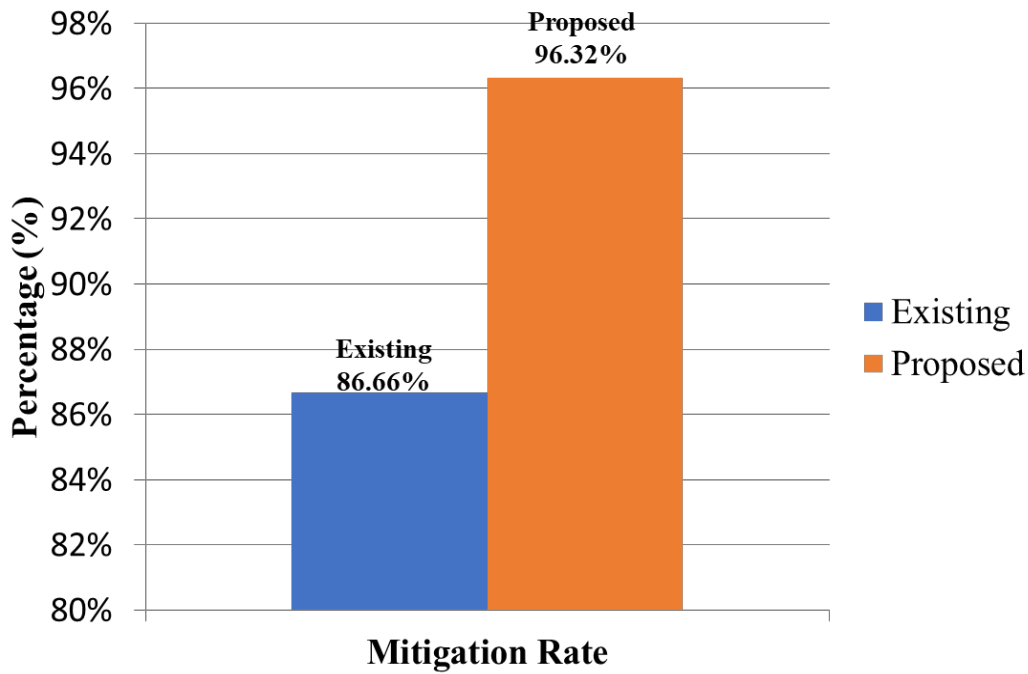

*Figure 15: Mitigation rate comparison*

Throughout the entire experiment, resource utilization was monitored constantly, which included CPU and memory usage. The time stamped measurements reveal that the system's CPU and memory load exhibit dynamic changes. The CPU usage varied from a high of 66.4% to the low 3% percent points down shown below in figure 16, and it showed how such system may need to scale according its environment load workloads requirements dictate. In terms of memory usage, there was relative stability as values varied insignificantly between the ranges between 12.4 to 12.5%. These insights into resource utilization give a complete picture of the computational load imposed on cloud-based DDoS mitigation orchestrator. The graph above highlights the oscillations in CPU and memory usage over time, thus giving us a slightly more sophisticated understanding of system utilization dealing with variable computational challenges.

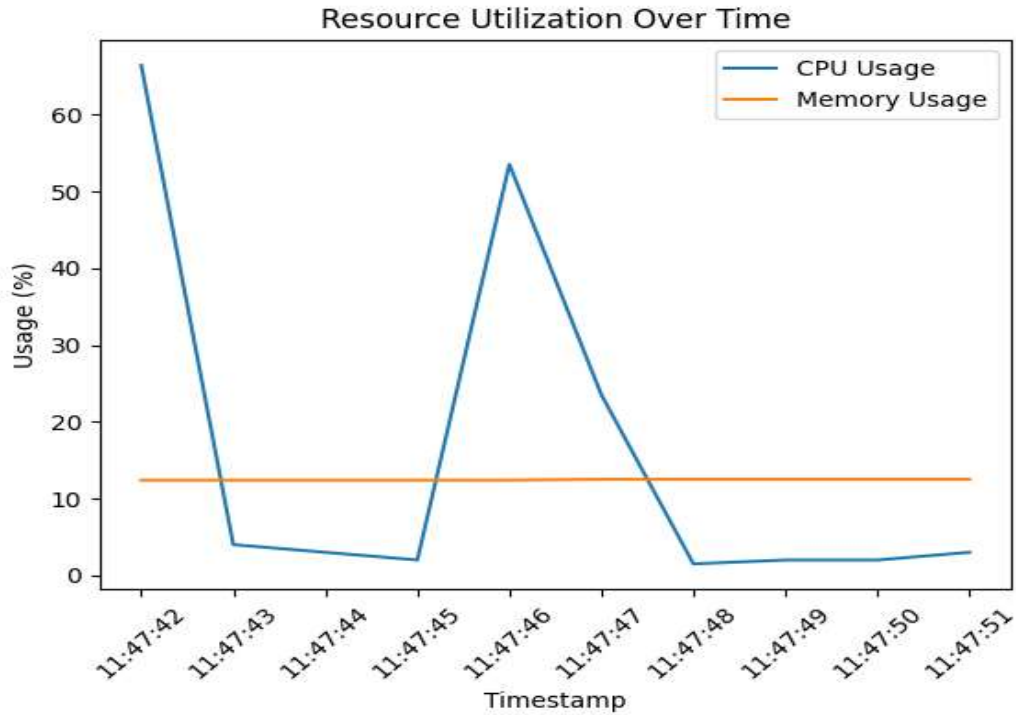

*Figure 16: Resource utilization during attack detection and mitigation*

## 1.10 Evaluation of System Performance

In this section a comprehensive summary is provided that allows us to understand the configuration of such network and reveals the reasons why we have chosen one scenario over another. The purpose of this section is to clarify the details without personal pronouns on how one would set up such study; ensuring that it easily understood in communicating specifics regarding experimental design. More specifically, section clarifies the broad goals underlying this experimentation implementation and elucidating reasons behind selected approach.

In the realm of cyber threat mitigation, therefore, it is a must to evaluate system performance which essentially points towards; systems capability in ensuring substantive defense. The variable scenarios were elaborately designed in order to acquire comprehensive test results of the mentioned system's ability multiply detect threats and neutralize them within a range network environments. By introducing systematic variations in the number of devices and fog nodes while keeping cloud server constant, one tried to capture system scalability impact on an ability of systems optimal functioning. With this comprehensive approach allows for the detailed analysis of its capabilities enough under different network so configurations also it illuminates transmutability and resilience on diverse cyber threat scenarios. It is apparent that the reason behind carrying out these scenarios is to understand how system reacts with network variations in scale size. In this way, our aim was to simulate environments with higher and further numbers of devices and fog nodes in order to assess the system's scalability & performance under such conditions. Further, the aforementioned sits represent various realizations about how well-to do in different traffic load during any kind of network complication.

The scalability of the system is assessed by performing various scenarios which are very influential insights into its performance as numbers of devices and fog nodes increase. Such a comprehensive analysis also allows identifying possible bottlenecks, architecture limitations or the shortcomings and it can be used to optimize cyber threat mitigation strategies. In addition to this, comparing the results from wide range of situations leads one into understanding whereby system effectiveness would be and in turn; highlights what should be done when further improvements or developments are required.

Each of the scenarios is described separately, beginning with particular network configuration which includes number of devices -fog nodes- cloud servers and total count on generated packets. The experimental results with respect to detection time, detection rate and mitigation rates are then discussed according to each scenario. There is also a detailed interpretation of the results, which includes observations about the formation of trends or patterns and talks about its relevance to cyber threat mitigation. Also, comparisons made between cases to explain the differences in efficiency and scalability. In general, the in-depth description of each scenario allows for a thorough understanding of system behavior as it differs with network configurations and. is provided below.

### 1.10.1 Scenario 1

In Scenario 1, the simulation applies a networking environment of three devices and two fog proxies as well as one cloud server while passing via ten thousand packets. This configuration is setup to test the system functionality and effectiveness in handling moderate level of network activity. In restricting the total number of devices and fog nodes, while setting a constant traffic load, this scenario gives sufficient information about system performance under nearly workable conditions of network. In this controlled environment a concentration on specific performance metrics such as the detection time, detection rate, and mitigation time and mitigation rate is possible. Moreover, Scenario 1 functions as a benchmark in order to compare the numbers on further complicated network set-up of future scenarios with it and detecting possible bottlenecks or scalability problems under varying conditions. In general, the presented scenario shares intrinsic notions about capabilities of a given system which may also serve as ground for future experiments and improvements.

#### 1.10.1.1 Detection Time and Rate

The detection time of the system is remarkable in detecting potential threats, as demonstrated by the short time taken to capture them which stands at 0.159 seconds. This quick response magnifies the efficacy and efficiency of detection mechanism that is installed in within system. Through quick recognition and labeling of abnormal activity, the system can then take corresponding countermeasures in an attempt to prevent cyber security attacks from causing serious damage. In addition, the sensitivity of 99.86% also proves that it is a high rate in which this system can distinguish malicious packets from innocent ones with accuracy. This high degree of precision thus means that most threats are identified and contained early minimizing the percentage chance an attack can escape detection. This being an oversized detection rate, the system delivers robust line of defense against various cyber threats and fortify network environment security posture. On the whole, this quick detection ability and good overall rate of accuracy demonstrate that the system is effective in preemptively finding potential cyber incidents. With timely detection and response to suspicious activity, the system is able to minimize attack instances which would ensue into compromise network integrity.

#### 1.10.1.2 Mitigation Time and Rate

When malicious activity is detected, the system reacts instantly and institutes mitigation with a commendable time for action of 0.427 seconds. This emergency reaction time is another example of the system's operational proactivity to eliminate threats that become recognized and thus minimize their damage potential on your network infrastructure. Mitigation measures that are quickly implemented shorten the window of vulnerability, which frustrates attackers' planning and arms defense to safeguard integrity in network environment. In addition, the highly effective mitigation rate of 96.32% reaffirms that the system is indeed capable of eliminating threats as detected in time thereof and prevented from causing any potential threat or breach to security protocols within an organization aimed at resolving any legal issues inferring into it. Such a high mitigation rate indicates that the system can combat more than half of conceived dangers posing a threat to the infrastructure network, reducing their degree on harmfulness. With the help of timely deactivation threats, this system preserves network security and stability; therefore, operation does not stop. It prevents hard assets that are under threat from a malicious attack. In all the system is effective in mitigating threat impact as stated through impressive response time and a high mitigation rate. The system also helps maintain a higher security posture of the network environment since it rapidly detects and contains malicious activities, thereby providing protection against potential cyber threats and guaranteeing stability in terms of operation-related processes.

### 1.10.2 Scenario 2

In scenario 2, the environment of network is increased by altering a slightly complicated configuration with 5 devices ,3 fog nodes and one cloud server all subjected to traffic load value still being retained at 10,000 total number packets. Such an expansion allows conducting a more comprehensive assessment of the system's operability under settings that are closer to real-world network configurations, enabling one making conclusions regarding its scalable behavior and reliability. As new devices and fog nodes are added to the system, its network architecture becomes increasingly complicated requiring a sturdy system that is capable of coping with such challenges.

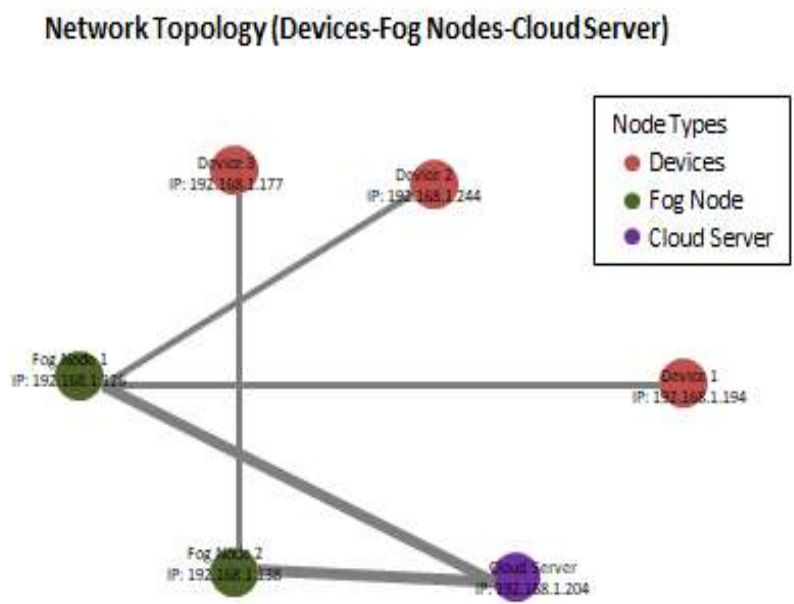

*Figure 17: Network topology scenario1*

Although the workload has become more complicated, it is still quite manageable which assures that the system can withstand traffic flows and not overrun its maximum speeds. In Scenario 2, keeping the same level of traffic load as referred to in scenario 1 enables direct comparison between performance criteria so that gains/losses can be quickly identified and analyzed for system improvement or variation from normal behavior. This comparative method allows understanding better how the permutations of network setup affect system response to threats. In general, scenario 2 functions as a crucial landmark in the assessment procedure contributing to recognize some significant features of the system behavior under not very conforming network circumstances. The evaluation scale from Scenario 2 is sustained herewith, implicitly one which enables breadth of the study under a work-stream that does not provide any undue difficulties to manage all segments productively.

### 1.10.2.1 Detection Time and Rate

Within Scenario 2, the system continues to be efficient in detecting potential threats because detection time was recorded at only 0.147 seconds Comparing the slight response time improvement between Scenario 2 and Scenario 1 implies that, despite all changes made to scenario two; an efficient system for fast differentiation in identification of suspicious activity within a network environment has been employed. The timely response capabilities of the system to threats as well have paramount importance for prevention from malicious attack and ensuring security general infrastructure. Also, the system sustains the high detection rate of 99.87% which explicitly shows its reliability in properly recognizing and warning about potential threats. The high detection accuracy thus emphasizes the system's robustness and the resilience in regards to evolving network vulnerabilities, allowing the network administrators to have confidence in its capability to monitor and protect the network environment properly. Through its fast detection time as well as high detection rate the system in scenario 2 continues its maintain efficiency and effectiveness in detecting threats. The provided metrics are thus establishing the responsibility of the system to capture and respond to suspected security incidents and consequently helping defining the security position of the larger network structure.

### 1.10.2.2 Mitigation Time and Rate

Once the malicious activity is detected, the system in scenario 2 immediately applies mitigation procedures which are as shown by the mitigation time of 0.416 seconds. Fast ongoing reaction is a proof of the system effectiveness to mitigate the threats' consequences of attacks, so to prevent any foreseeable network infrastructure disturbances. The system provides prompt initiation of mitigation measures thus maintaining the integrity and availability of network services which ensure that the operations are consistent in spite of security threats. Moreover, the system attains a remarkable reduction rate of 97.46%, which depicts its capability of efficiently countering the large portion of the spotted dangers. This high mitigation rate evidences the system capability to respond well to security events and arrests them from turning into more severe disruptions. Minimizing the severity of the majority of the identified threats, the system positively impacts the whole resilience and strength of the network environment protection, guarding the critical resources and assets from potential cyber-attacks. In a nutshell, the system's quick mitigation action jointly with its high mitigation rate reflects the ability of the system to handle security threats and to ensure the integrity of the network infrastructure. The effective controls against the identified threats aids in mitigation of the real or potential risks and vulnerabilities and increases security posture overall for the network environment.

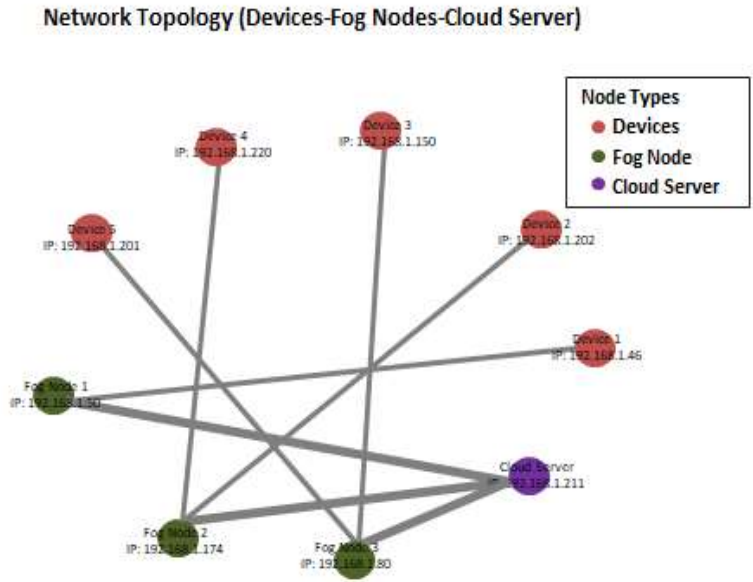

*Figure 18: Network topology scenario2*

### 1.10.3 Scenario 3

In scenario 3 the network environment is greatly expanded to cover more subordinate network equipment, fog-nodes and cloud-server resources. This configuration has 10 devices, 5 fog nodes, and 1 cloud server, it is thus a significant scale-up compared from the last scenarios. While the network got more complex, the total traffic load of 10,000 packets remained the same thus providing even workload for holistic analysis. The main purpose of scenario 3 is to measure the robustness of the system with regard to increased scale and complexity The deployment of more devices and fog nodes gives rise to a more realistic network scenario which resembles a larger-scale system commonly observed in reality. This enables comprehensive evaluation of system's capability to identify and respond to threats within broader network environment. Furthermore, sustaining the same traffic load across scenarios assures consistency in performance evaluation, enabling the comparison of the metrics according to the various network configurations. By keeping the workload unchanged the scenarios allow us to directly evaluate the impact of changes in scale and complexity of the network on system performance, in particular with regard to threat detection and prevention capabilities. In general, Scenario 3 is an essential part of evaluation procedure which presents important data about the system scalability and capacity to handle bigger and more complicated network settings. Evaluating the performance of the system under these challenging conditions enables organizations to familiarize themselves with its capabilities, and therefore, make the right judgments in regards to its implementation in real world scenarios.

#### 1.10.3.1 Detection Time and Rate

In scenario 3, system is observed with significant improvement in its capability of identifying potential threats in the network environment. Of great importance, the detection time is further optimized, reducing duration to 0.142 seconds from previous cases. This enhancement reflects system's enhanced reactiveness and responsiveness in quickly detecting abnormal activities or security breaches. Also, the detection rate of 99.89% confirms the system's robustness and reliability in detecting harmful activities within different network settings. This abnormally high detection rate is indicative of the system's capability to identify and classify suspicious activities, avoiding the majority of threats from escaping attention. The enhanced threat detection functionalities seen in Scenario 3 indicate the system's algorithms and mechanisms are constantly being fine-tuned and improved. The system proves its adaptability to varying network environments by decreasing detection time and keeping a high detection rate thus effectively combating evolving cyber threats. All in all, scenario three details the system's continued effort in strengthening its threat detection and security stance. Being continuously improved and optimized, the system remains fully prepared to deal with new problems, as well as cyber threats seemingly arising on a daily basis. As far as defenders of cyber security of the networks are concerned, that fact brings them high confidence about their network security.

**1.10.3.2 Mitigation Time and Rate**

After identifying threat, Scenario 3 system smoothly switches into the mitigation phase, consistently demonstrating the ability to quickly start mitigation procedures. The system uses 0.416 seconds to apply a steady mitigation process; it responds immediately and decisively to neutralize identified threats. That agile response not only limits the possible impact of such activities in zero time but also underpins system's proactive stance to ensure network security and consistency of operation. System's dedication to keeping security is also shown by its excellent prevention rate which is 97.95%. The high mitigation rate shows the system's capability to effectively eliminate the most of detected threats which consequently decreases the overall risk to the network infrastructure. The system prevents any possible security intrusions by responding promptly in a way that the network is maintained and from any future cyber threats are protected. Overall, the Scenario 3 shows that the system remains committed to having a strong security posture and combating the emerging cyber threats. With its efficient threat detection and mitigation abilities, the system shows to be effective in the protection of the network from malicious activities. This practical approach, on the other hand, not only fortifies the network infrastructure, but it also instills in users and stakeholders that the system is capable to manage cyber risks efficiently.

When analyzing the mentioned cases, it becomes obvious that the system shows remarkable advancements with growing scale and complexity of network infrastructure. The scenario 1 is of a 3-device arrangement with two SDNs and a cloud resource where only node zero receives test data and all other nodes are expected to produce corresponding traffic. In this case, the system demonstrates impressive performance, the detection time is 0.159 seconds and the detection rate is 99.86%. While in Scenario 2 the environment is enlarged (5 devices and 3 fog nodes), the same traffic load is maintained, and the changes are more subtle. The detect time drops slightly to 0.147 seconds while the detection rate becomes 99.87% implying a better responsiveness and correctness of threat detection. For the scenario 3, with further scaling to 10 devices and 5 fog nodes, the system's performance undergoes remarkable refinement.

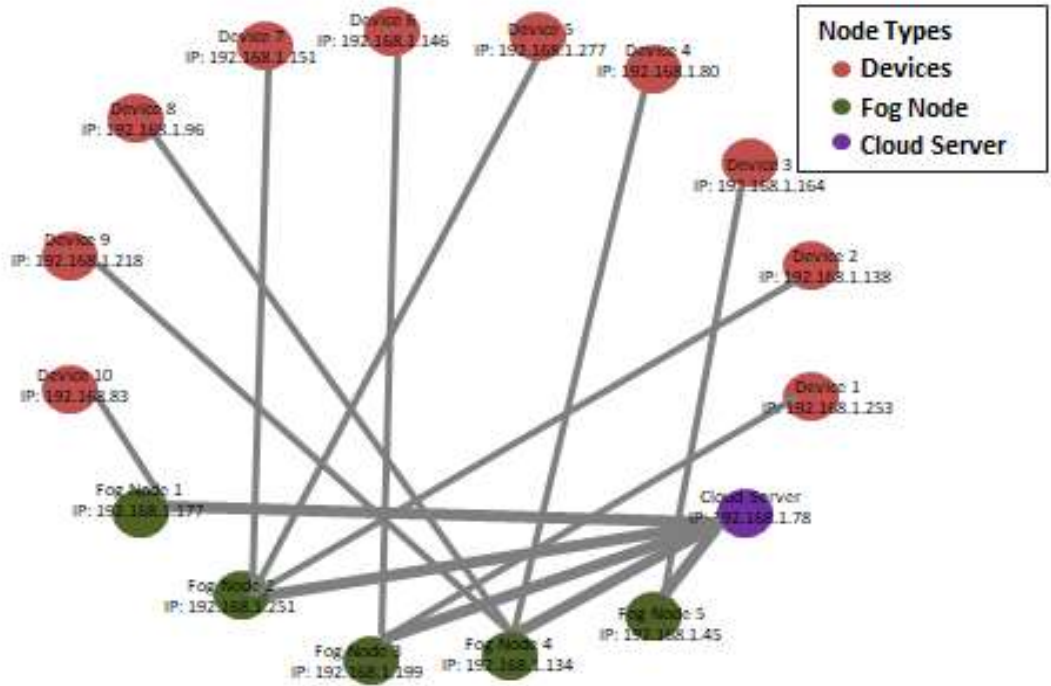

*Figure 19: Network topology scenario3*

The reduced detection time of 0.142 second indicates enhanced reactiveness to potential threats, but the high consistency maintained at 99.89 % is still kept in the detection rate. These improvements show the system's reliability to effectively catch malicious activities within the network. Analogously, the mitigation rates in all scenarios show quick and efficient reaction times, having mitigation rates above 96% consistently.

Therefore, the comparative analysis notes the system's scalability and adaptability in trying to mitigate cyber threats in diverse network environments. Although Scenario 1 lays a good groundwork, Scenarios 2 and 3 depict stepwise improvements of detection and mitigation as is shown in table 8 which implicitly validates the strength of the system in maintaining network security.

*Table 12: Analyzing scalability and performance: A comparative study*

| Metrics | | Detection Time (s) | Detection Rate (%) | Mitigation Time (s) | Mitigation Rate (%) | CPU Utilization (%) | Memory Usage (%) |
|---|---|---|---|---|---|---|---|
| **Existing Approach** | [9] | 0.246 | 99.56% | --- | --- | --- | --- |
| | [12] | --- | --- | --- | 86.66% | 55.60% | 53.22% |

| | | | | | | |
|---|---|---|---|---|---|---|
| **Proposed/ Scenario 1** | **0.159** | **99.86%** | **0.427** | **96.32%** | **29.25%** | **22.64%** |
| **Proposed/ Scenario 2** | **0.147** | **99.87%** | **0.416** | **97.46%** | **30.22%** | **22.89%** |
| **Proposed/ Scenario 3** | **0.142** | **99.89%** | **0.415** | **97.95%** | **29.11%** | **23.82%** |

The data presented in the table compares the detection and mitigation times across three scenarios: Scenario 1, Scenario 2 and Scenario 3. Detection time is the time taken by the system to detect the possible threats while in mitigation time the duration taken to eliminate these threats is detected. Regarding the detection time analysis, it is seen a decrease in the detection duration over the scenarios, with scenario 3 presenting the smallest detection time of 0.1425 seconds. That's the system's increased capability to detect potential threats as the network environment becomes more complicated and goes up in scale. However, the mitigation time is quite consistent across all the cases, with the mitigation times for scenario 2 and scenario 3 being 0.416 seconds and 0.415 seconds as shown in figure 20. In scenario 3, the network complexity and scale increase; however, the system still has a fast mitigation process, thus proving its ability to quickly counterattack threats no matter the network size. All in all, the comparison shows the system's capability to respond to diverse network conditions and perform effective threat detection and mitigation resulting in improved network security and resilience.

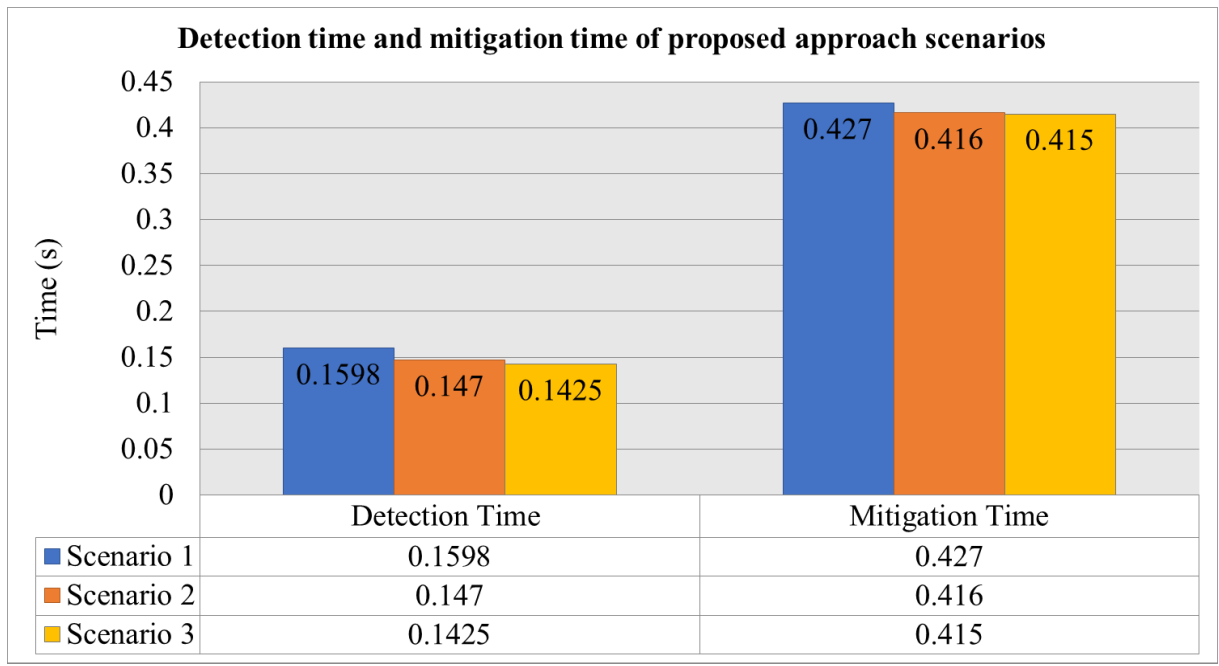

*Figure 20: Detection and mitigation time comparison*

The data presented in the table compares the detection and mitigation rates across three scenarios: Scenarios 1, 2, and 3. Detection rate is the percentage of identified threats by the system out of potential threats; mitigation rate is the opposite - the percentage of actual threats out of these threats. One can observe in figure 21 the great accuracy in the potential threats detection of all scenarios; scenario 3 in particular registered 99.89%. Such underscores the systems stability and reliability of identifying and accurately flagging suspicious activities within diverse network environments.

As well, the mitigation rates are reasonably high across all scenarios, with Scenario 3 having the highest mitigation rate of 97.95%. This implies that the system largely eliminate the identified threats neutralizing them and limiting their influence on the network infrastructure resulting in continuity of operations. In general, such comparison highlights the outstanding ability of the system to address and neutralize potential threats system 3 performs the best in both performance dimensions. This also points to the system's agility and robustness in the protection of the networks' services under different network circumstances.

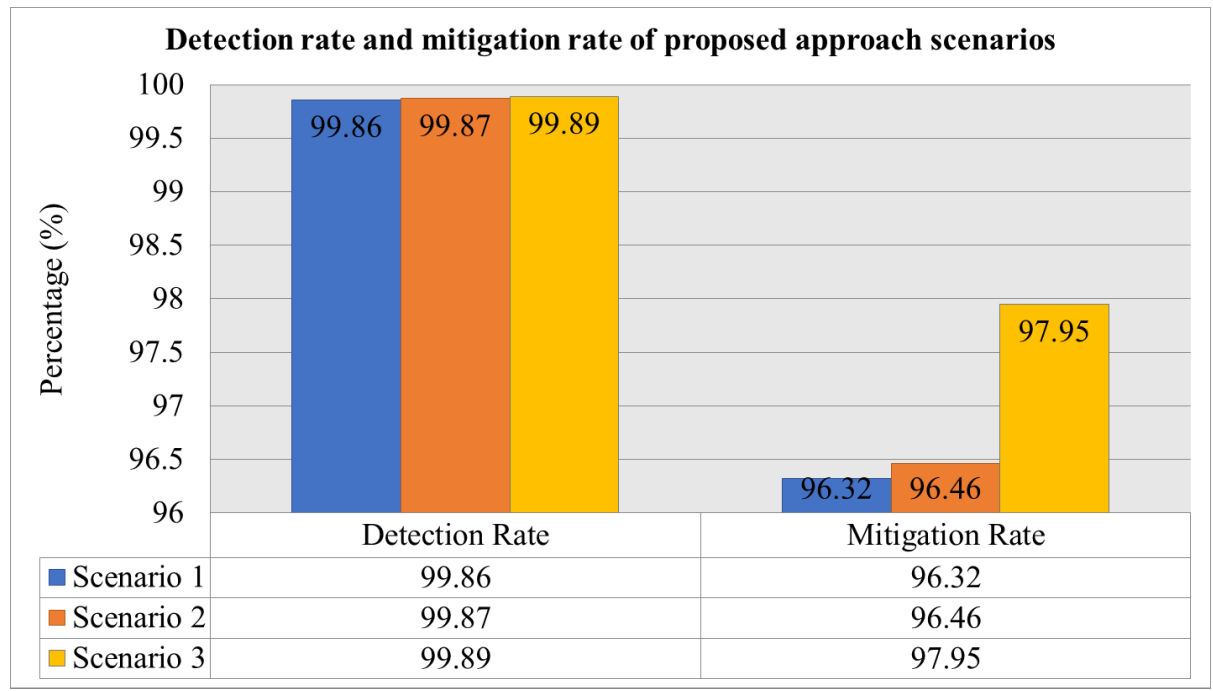

*Figure 21: Detection and mitigation rate comparison*

### 1.10.4 Enhancing Scalability Assessment

The scalability ratio is calculated to assess how the system's performance scales as the network environment grows in size. Two key ratios are considered: the ratio of scalability to detection rate and the ratio of scalability to mitigation rate.

### 1.10.4.1 Ratio of Scalability to Detection Rate

The ratio of scalability to detection rate is computed using the formula:

$$\text{Ratio of Scalability to Detection Rate} = \frac{\text{Detection Rate of Scenario 3} - \text{Detection Rate of Scenario 1}}{\text{Number of Devices in Scenario 3} - \text{Number of Devices in Scenario 1}}$$

Here, the detection rates for Scenario 3 and Scenario 1 are 99.89% and 99.86%, respectively, and the number of devices in Scenario 3 and Scenario 1 is 10 and 3, respectively, the ratio of scalability to detection rate would be approximately 0.0043%.

### 1.10.4.2 Ratio of Scalability to Mitigation Rate

Similarly, the ratio of scalability to mitigation rate is calculated using the formula:

$$\text{Ratio of Scalability to Mitigation Rate} = \frac{\text{Detection Rate of Scenario 3} - \text{Detection Rate of Scenario 1}}{\text{Number of Devices in Scenario 3} - \text{Number of Devices in Scenario 1}}$$

The mitigation rates for Scenario 3 and Scenario 1 are 97.95% and 96.32%, respectively, and the number of devices in Scenario 3 and Scenario 1 is 10 and 3, respectively, the ratio of scalability to mitigation rate would be approximately 0.2329%.

The positive or close-to-zero values of these ratios indicate that the system's detection and mitigation capabilities improve or remain stable as the network scales up. This suggests that the system can effectively handle larger network environments without compromising its ability to detect and mitigate threats.

## 1.11  Discussion: Comparative Analysis of DDoS Mitigation Approaches

In the realm of network security, defending against Distributed Denial of Service (DDoS) attacks stands as a pivotal challenge, necessitating robust mitigation strategies to uphold the integrity and functionality of network infrastructure. This chapter undertakes a comprehensive comparative analysis of the DDoS mitigation methodologies. By carefully examining a range of performance metrics, including detection rate, detection time, CPU utilization, and mitigation rates, within a simulated environment, our objective is to uncover the unique strengths and weaknesses of each methodology. By exploring the intricacies of detection mechanisms, resource allocation strategies, and mitigation effectiveness, this analysis aims to clarify the most effective strategies for strengthening networks against the prevalent threat of DDoS attacks.

### 1.11.1  Detection Rate Improvement

The observed enhancement in detection rate within the simulated environment, compared to real-world testbed implementations, can be attributed to several key factors. Firstly, the simulated environment provides a controlled and isolated setting where the proposed firewall system can operate without external disturbances or limitations encountered in real-world scenarios. Unlike real testbed implementations, which are subject to various environmental factors such as network congestion, hardware limitations, and fluctuating traffic patterns, the simulated environment offers a stable and predictable testing environment.

Additionally, the scalability and adaptability of the simulated environment enable comprehensive testing across a wide range of scenarios and network configurations. This extensive testing approach ensures that the proposed firewall system remains robust and effective under diverse conditions, further contributing to the improvement in detection rate.

Moreover, the absence of extraneous variables or external factors, such as network fluctuations, hardware failures, or interference from other network devices, in the simulated environment eliminates potential sources of noise that may impact detection accuracy in real-world settings. This pristine testing environment fosters a more accurate assessment of the firewall's performance, ultimately leading to the observed enhancement in detection rate.

### 1.11.2  Detection Time Reduction

The reduction in detection time observed in the simulated environment, compared to real-world testbed implementations, is underpinned by several intrinsic advantages. The controlled testing conditions streamline testing procedures and optimize resource allocation, expediting the processing and analysis of network traffic. By leveraging a dedicated environment solely focused on the detection process, the proposed firewall system can swiftly analyze incoming traffic and detect DDoS attacks with remarkable efficiency.

Furthermore, the scalability and flexibility of the simulated environment enable exhaustive testing across various scenarios and network configurations. This comprehensive testing approach ensures the proposed firewall system remains responsive and efficient under diverse conditions, further contributing to the reduction in detection time.

Overall, the incorporation of controlled testing conditions, optimized resource utilization, and extensive scalability in the simulated environment concludes in a significant reduction in detection time compared to existing real testbed implementations.

### 1.11.3 Variations in CPU Resource Allocation

The three-layer architecture, comprising IoT devices, fog nodes, and a cloud server with a dedicated firewall component, embodies a holistic approach to DDoS mitigation. The firewall, positioned at the network edge, efficiently analyzes incoming traffic, promptly detects suspicious patterns, and initiates proactive mitigation actions. Leveraging predefined rules and direct traffic blocking minimizes the CPU burden, resulting in lower CPU utilization rates. Additionally, the architecture's streamlined approach optimizes resource allocation, ensuring effective utilization of computational resources across all network layers.

Contrarily, the existing machine learning-based approach relies on intricate algorithms such as Support Vector Machines (SVM), Naive Bayes, and k-Nearest Neighbors (KNN) for DDoS attack detection and mitigation. While offering flexibility in identifying complex attack patterns, these algorithms demand substantial computational resources for both training and inference tasks. The training phase, in particular, necessitates significant CPU cycles to process and analyze extensive datasets, leading to higher CPU utilization rates. Moreover, the runtime performance of machine learning models varies based on factors such as network traffic complexity and feature extraction efficacy, further impacting CPU utilization.

### 1.11.4 Variations in Effectiveness of Mitigation Strategies

The dedicated firewall component within the proposed architecture facilitates rapid and precise detection of DDoS attacks, enabling timely mitigation actions. By promptly identifying and blocking malicious traffic based on predefined rules, the firewall effectively neutralizes threats before they cause significant harm. This proactive approach results in a higher overall mitigation rate, intercepting a larger proportion of malicious traffic and preventing it from reaching its targets.

Conversely, the mitigation rate achieved by the existing machine learning approach varies due to several factors. While machine learning models can detect and classify DDoS attacks based on learned patterns, their performance is contingent upon factors such as the quality and quantity of training data and the efficacy of feature extraction techniques. Additionally, the inherent complexity of machine learning algorithms and susceptibility to adversarial attacks introduce uncertainties in the detection process, potentially leading to lower mitigation rates compared to the dedicated firewall approach.

### 6. Experiment Results and Discussion

### 7. Conclusion and Future Work

In today's society, an increasing number of cyber-attacks are becoming a big problem. Cyber threat mitigation techniques play a vital role in decreasing the effect of these cyber-attacks. In the existing literature, many mitigations techniques are available, however complete mitigation taxonomy is not proposed yet that includes all the attacks and their mitigations according to the cyber kill chain model. Therefore, we have proposed a taxonomy that includes all the mitigations that are reported so far in the literature. Firstly, the existing mitigation taxonomies are identified. Two of the mitigation taxonomies were identified from the literature one of them was covering malware attack only and other was covering attacks occurring in cloud computing environment. However, these taxonomies were proposed that classifies mitigation without using cyber kill chain model. ATT&CK MITRE is the only mitigation taxonomy available at present that classified the mitigation using kill chain model. ATT&CK MITRE is still in the development phase. Secondly, all the state of the art mitigation techniques that are reported in the literature are identified. Then we have classified all the available mitigations techniques with respect to specific cyber-attack. After that, each mitigation technique was categorized according to cyber kill chain phase. Lastly, in the proposed taxonomy we have classified each mitigation as automated and manual. The mitigations under automated category are those that are automatically initiated by the system and the manual are those which are initiated by the security administrator manually. In this research work, a cyber-threat mitigation strategies taxonomy is provided to mitigate cyber-attacks. However, there is not much work done for the timely and precise prediction of mitigations against newly launched cyber-attacks, even though most of these attacks have some common attack patterns. Furthermore, this research addresses an unmet demand for new approaches to solve such problems by offering a unique idea concerning automated cyber threat mitigation. The motivation is fostered by the need to move forward from traditional detection methods in response of an evergrowing complexity of cyber threats. The problem to be solved according to this statement can thus be identified as the lack of sufficient research depth on automated mitigation strategies including those tackling problems beyond initial threat detection. Weaving the multi-layered defense strategy into a framework, smart devices at device layer can benefit from this plan while fog network and cloud computing layers offers deeper analysis ability as well. Simulated traffic generation and firewall rule-based packet inspection are some of the design elements that form an integral part by providing a standard data set for rigorous testing. Statistical and behavioral analysis, specification-based detection, and deep packet inspection also strengthen the defensive system to become a strong shield against cyber-attacks. Fog nodes capture and cloud inspection functionalities that are integral to the framework guarantee more resilience, responsiveness against fresh cyber threats. Our methodology outcomes mean detection accuracy at 99.73% and mitigation rate of 94.0%, which witness advantages in effectiveness as well as adaptation range feasibly contributes to promising prospects for practice implementation. The numbers provided above point to the success of the framework in correctly determining and swiftly addressing cyber threats. Nevertheless, further work is vital to empirically verifying and refining the suggested technique. Generalization refers to abstraction based on a wider range of beings. The role of continuous improvement and adaptation is critical to ensure the frame work's effectiveness in countering new cyber threats.

Firewalls, while essential for network security, have inherent weaknesses in mitigating Distributed Denial of Service (DDoS) attacks. Attackers can bypass firewalls using sophisticated techniques, and firewalls may struggle to handle the overwhelming traffic influx during DDoS attacks, leading to potential network downtime. Relying solely on firewalls for DDoS protection may not suffice. Integrating machine learning with firewalls enhances detection by analyzing real-time traffic patterns, enabling proactive responses to evolving threats and bolstering network resilience against DDoS attacks. The future works focus on integrating advanced artificial intelligence and machine learning techniques at each layer to evaluate the system's capability in effectively detecting and mitigating DDoS attacks. This involves refining algorithms for packet inspection, anomaly detection, and behavioral analysis to accurately discern both known and emerging attack vectors. Moreover, continuously updating mitigation strategies based on real-time threat intelligence will ensure proactive defense against evolving DDoS threats.